\documentclass[12pt,a4]{article}

\usepackage{float}
\usepackage{latexsym}
\usepackage{psfig}
\usepackage{rotating}
\usepackage{graphics}
\usepackage{epsf}
\usepackage{epsfig}
\usepackage{float}
\usepackage{fancyhdr}

\newcommand{\rp}{\mbox{$\not \hspace{-0.15cm} R_p~$}}

\begin{document}

\begin{center}
{\Large \bf Search For New Particles at HERA}\\

\vspace{4mm}

Mireille Schneider\\
CPPM, 163, avenue de Luminy, case 907,\\
13288 Marseille cedex 9, France

\end{center}

\begin{abstract}

Recent results on searches for physics
beyond the Standard Model obtained by the H1 and ZEUS 
experiments are reported here. After a brief introduction to the HERA collider,
indirect searches for contact interactions and extra-dimensions
are presented as well as direct searches for new physics including
leptoquarks, lepton-flavour violation, squarks produced by R-parity violation and
excited fermions. New results from isolated lepton events and single top searches
are also presented. Finally the future prospects of HERA-2 are shown.

\end{abstract}

\section{Introduction}

HERA is an $ep$ collider which has especially high
sensitivity to new particles coupling to
lepton-quark pairs.
In 1994-97 HERA collided 27.5 GeV positrons on 820 GeV protons. In 1998 the
proton energy was raised to 920 GeV increasing the
center-of-mass energy $\sqrt{s}$
from 300 GeV to 318 GeV. In 1998 and in the first months
of 1999, HERA ran with electrons. In May 1999 HERA switched back to $e^+p$
collisions.

The three main colliding periods as well as the corresponding luminosities 
for each experiment are summarized in table \ref{tab:lumit}.

\bf
\begin{table}[hhhh]
\begin{center}
\begin{tabular}{|c|c|c|c|c|}
\hline
Year & Collision & $\sqrt{s}$(GeV) & H1 & ZEUS \\
\hline
94-97 & $e^+p$ & 300 
& $ 36~\mbox{pb}^{-1}$ & $ 48~\mbox{pb}^{-1}$\\
\hline
98-99 & $e^-p$ & 318 
& $ 14~\mbox{pb}^{-1}$ & $ 16~\mbox{pb}^{-1}$\\
\hline
99-00 & $e^+p$ & 318 
& $ 65~\mbox{pb}^{-1}$ & $ 64~\mbox{pb}^{-1}$\\
\hline
\end{tabular}
\end{center}
\vspace{-0.5cm}
\caption{ \it Luminosities collected by H1 and ZEUS for each colliding period.}
\label{tab:lumit}
\end{table}
\rm

Recent results on searches for new particles based on different models 
beyond the SM are reported here, as well as searches for 
rare event topologies with high $P_T$ leptons and missing momentum.

Several models have been tested analysing the full HERA-1 data corresponding
to an integrated luminosity of 115 and 128 $\mbox{pb}^{-1}$
for H1 and ZEUS  respectively. 

\section{Contact Interactions}

This indirect search for new processes allows to test the effects of 
new physics implying the exchange of particles with a mass higher than
the center-of-mass energy $s$. An effective mass scale $\Lambda$ is introduced:
$\Lambda~\gg~\sqrt{s}$.

Different models beyond the Standard Model
(SM) can thus be tested in the frame of
contact interactions: for instance,
in compositeness models, limits on the mass scale $\Lambda$ can be set.
Limits on high mass leptoquarks can also be determined via the ratio 
$M/\lambda$ where $\lambda$ is the coupling between leptoquarks and SM particles.
Finally results from contact interactions allow also to put limits 
on the radius of the quark $R_q$.

Contact interactions are represented by lagrangian (\ref{eq:cilag}) which is added
to the SM lagrangian.

\begin{equation} \label{eq:cilag}
{\cal L}_{CI}=\sum_{a,b=L,R} \eta_{ab}^q(\overline{e_a} \gamma_\mu e_a)
(\overline{q_b} \gamma_\mu q_b)
\end{equation}

$\eta_{ab}^q=\epsilon \frac{g^2}{(\Lambda_{ab}^q)^2}$ are model-dependant
coefficients depending on the quark flavour $q$ and the quark and electron 
chiralities $a$ and $b$.
Different combinations of these coefficients lead to different
contact interaction models. $\epsilon=\pm 1$ is the sign of the interference
with respect to the SM, $\Lambda_{ab}^q$ is the effective mass scale, and
$g$ is the coupling constant, conventionally fixed so that $g^2=4\pi$.

Contact interactions are searched for by analysing high $P_T$ samples
of deep inelasting scattering candidates (DIS samples). Deviations from the SM
in the distributions of usual kinematic variables of the DIS candidates
(e.g. the virtuality $Q^2$ of the exchanged photon) are looked for.

\begin{figure}[hhhh] 
\vspace{-1.cm}
$$
\epsfig{figure=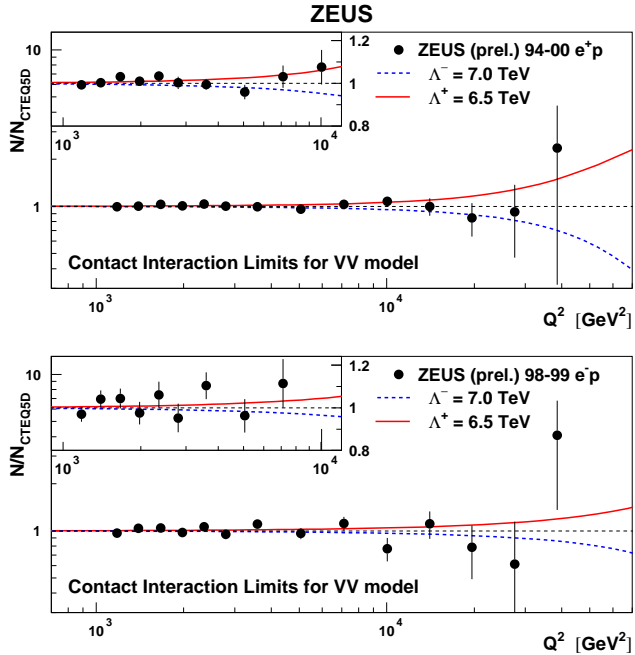,height=10.cm}
$$
\vspace{-1.5cm}
\caption{ \it Ratio between the experimental and SM cross sections
as a function of $Q^2$ according to the VV compositeness model.
New physics interferes constructively (solid curve) or destructively (dashed curve)
with the SM DIS.}
\label{fig:exferm}
\end{figure}

As an example, 
figure \ref{fig:exferm} shows a fit of the measured neutral
current (NC) $Q^2$ distribution
to the VV compositeness model 
(this model assumes the same couplings $\eta_{ab}^q$ for all particle states),
in both cases where new physics interferes constructively or destructively
with the SM DIS processes.
No deviation at high $Q^2$ is observed, therefore limits are deduced.

\begin{figure}[hhhh] 
$$
\epsfig{figure=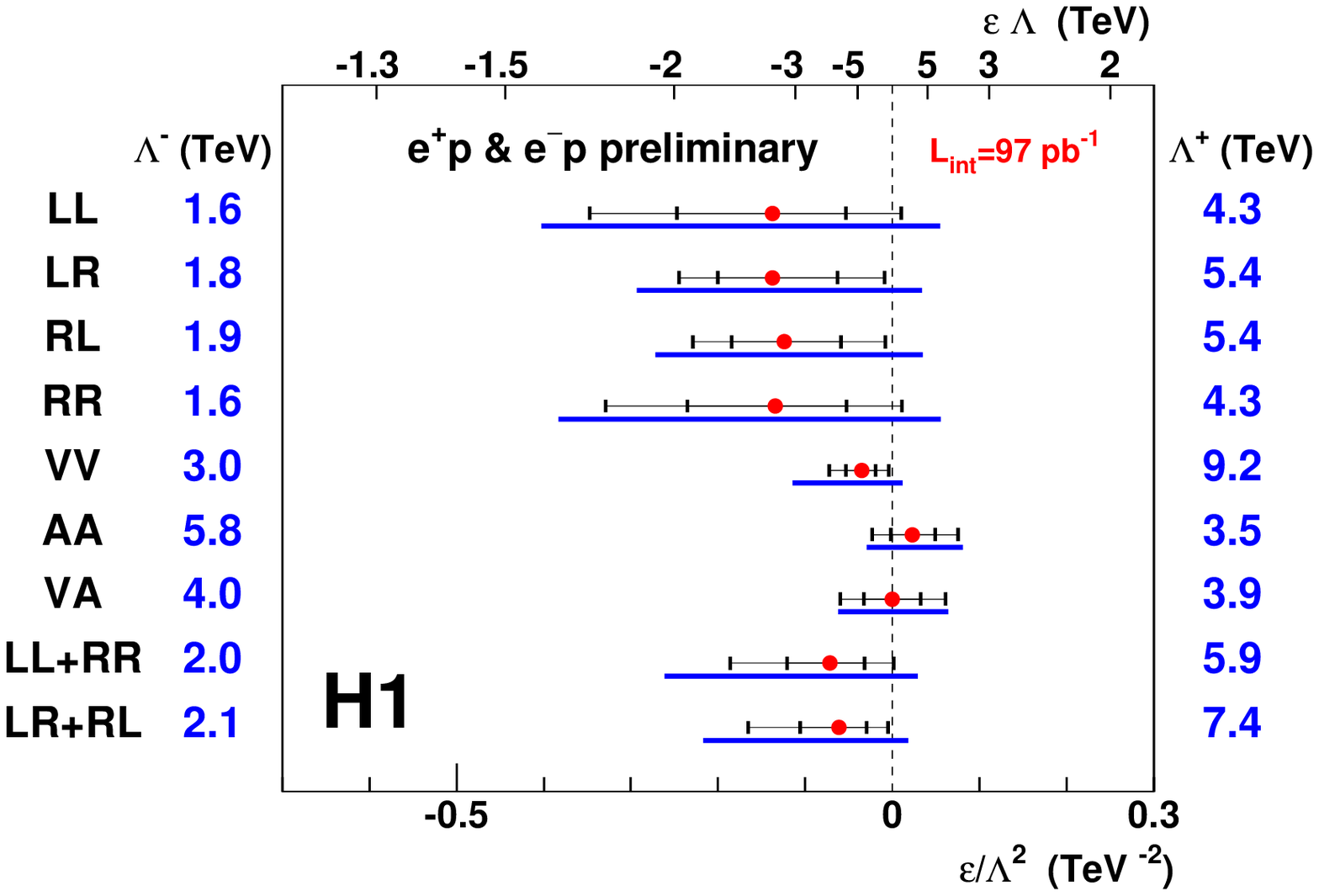,height=6.8cm}
\hspace{-0.5cm}
\epsfig{figure=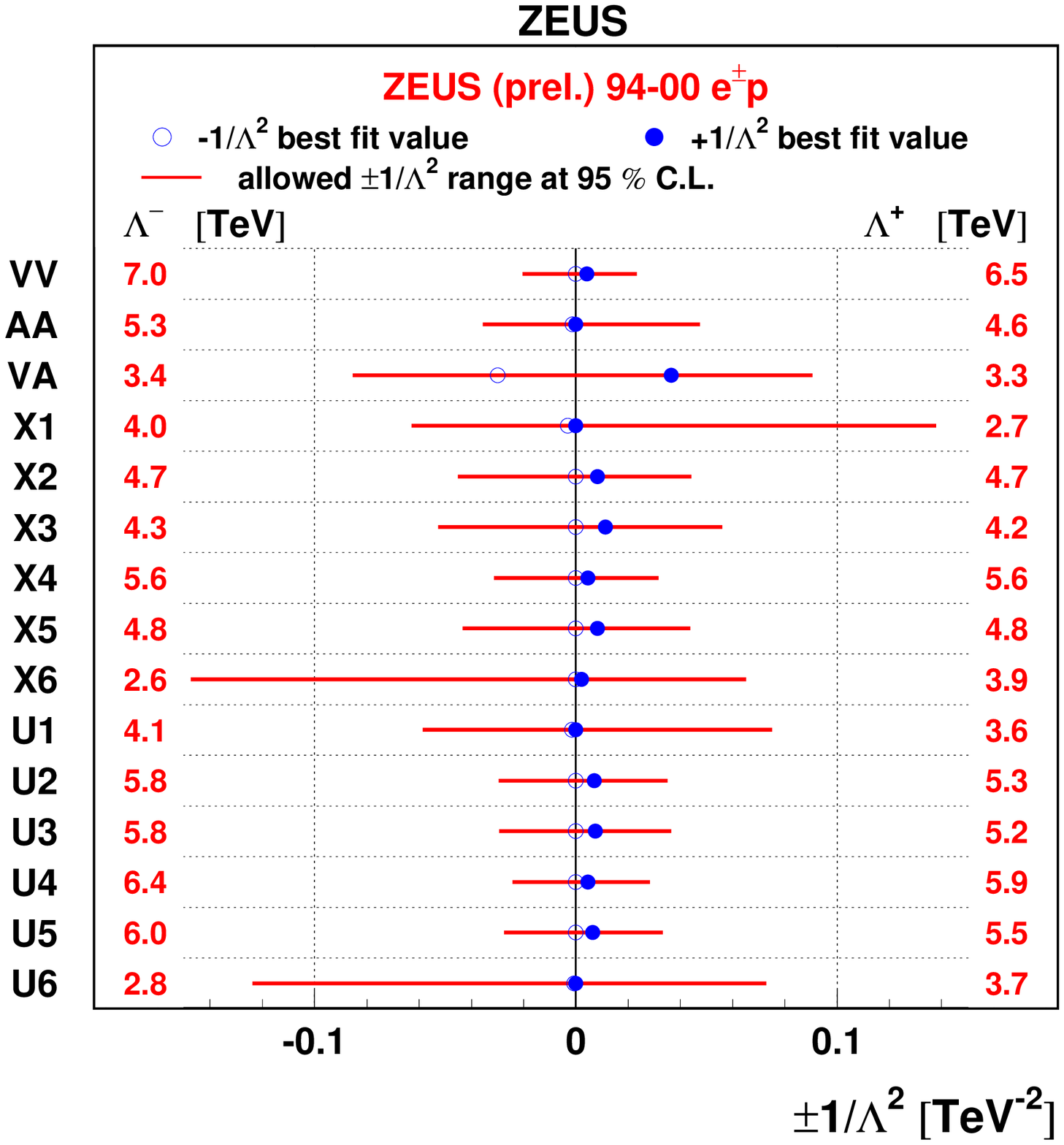,height=6.8cm}
$$
\vspace{-0.5cm}
\caption{ \it Limits on $\epsilon/\Lambda^2$ for different
compositeness models obtained by H1 and ZEUS. The thick lines in
first (second) figure show the H1 (ZEUS) limits on $\epsilon/\Lambda^2$
analysing 97 (128) $\mbox{pb}^{-1}$ of $e^\pm p$ data.
Values outside these regions are excluded at $95\%$ CL.}
\label{fig:cigen}
\end{figure}

Figure \ref{fig:cigen} shows the H1 and ZEUS exclusion limits on the ratio
$\epsilon/\Lambda^2$ for general compositeness models
\cite{h1cipap}. H1 and ZEUS data are in good agreement with the SM
expectation.

The resulting limits on the effective scale $\Lambda$ are summarized in
table \ref{tab:cilim}.

\bf
\begin{table}[hhhh]
\begin{center}
\begin{tabular}{|c|c|}
\hline
H1 & $\Lambda>1.6$ - $9.2~\mbox{TeV}$ \\
\hline
ZEUS & $\Lambda>2.6$ - $7.0~\mbox{TeV}$\\
\hline
\end{tabular}
\end{center}
\vspace{-0.5cm}
\caption{ \it Limits on the effective mass scale $\Lambda$ obtained by H1 and ZEUS.
The ranges correspond to the different compositeness models.}
\label{tab:cilim}
\end{table}
\rm

Measurements of high $Q^2$ DIS can also be used to set limits on the 
radius of the quark 
$R_q$: the measurement of the spatial charge distribution of the quark is performed
using the classical form factor approximation. In this approximation, the 
structure of the quark leads to the cross section deviation from the SM
presented in equation (\ref{eq:dev}).

\begin{equation} \label{eq:dev}
\frac{d\sigma}{dQ^2}=\frac{d\sigma^{SM}}{dQ^2}\left[ 1-\frac{R^2_e}
{6}Q^2\right]^2 . \left[ 1-\frac{R^2_q}{6}Q^2\right]^2
\end{equation}

$R_e$ and $R_q$ are the root of the mean square radius of the electroweak 
charge of the electron and the quark respectively.

Assuming that electrons are point-like particles, an upper limit of $0.73.10^{-18}$
m is obtained for the quark radius from a fit to the ZEUS data \cite{zeusci}.

\section{Extra-Dimensions}

The Large Extra-Dimensions (LED) model,
proposed by Arkani-Hamed, Dimopoulos and Dvali assumes a space-time
containing $4+n$ dimensions \cite{led}:
the SM particles including the strong
and electroweak bosons are supposed to be confined in a 4D world while 
gravitons can propagate in $n$ compactified extra-dimensions.

The effective coupling strength describing the interaction between the 
gravitons and
SM particles is $1/M_s$, where $M_s$ is the fundamental Planck mass
scale in the full space.
If extra-dimensions are large enough, then $M_s$ may be in the TeV range. 
The contribution of graviton exchange to $eq$ scattering can be 
described by a contact interaction. Therefore the model is tested using 
NC samples and fitting the cross section to a formula including 
the graviton exchange, considering $M_s$ as a free parameter.

The corresponding limits on the Planck mass scale $M_s$ 
for both experiments are summarized in table \ref{tab:ledt} 
\cite{zeusci,h1lqichep}.

\bf
\begin{table}[hhhh]
\begin{center}
\begin{tabular}{|c|c|c|}
\hline
 & Negative Coupling & Positive Coupling\\
\hline
H1 (96 $\mbox{pb}^{-1}$) & $M_s^->0.93$ TeV & $M_s^+>0.63$ TeV \\
\hline
ZEUS (128 $\mbox{pb}^{-1}$) & $M_s^->0.82$ TeV & $M_s^+>0.81$ TeV\\
\hline
\end{tabular}
\end{center}
\vspace{-0.5cm}
\caption{ \it Limits on the Planck mass scale $M_s$ with H1 and ZEUS 
detectors.}
\label{tab:ledt}
\end{table}
\rm

\section{Leptoquarks}

\begin{figure}[hhhh] 
$$
\epsfig{figure=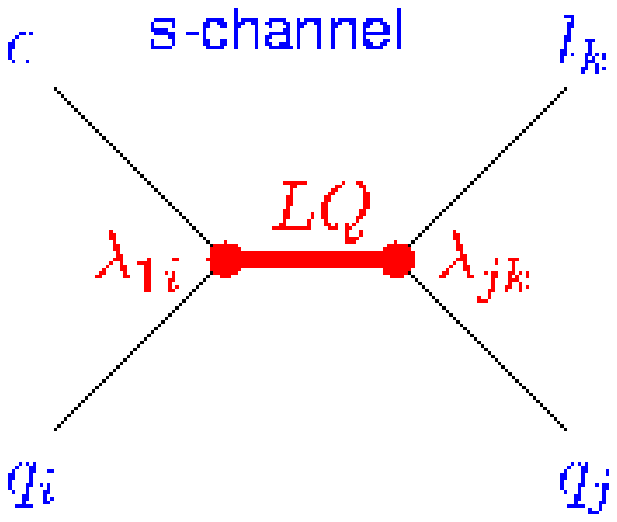,height=3.cm}
\epsfig{figure=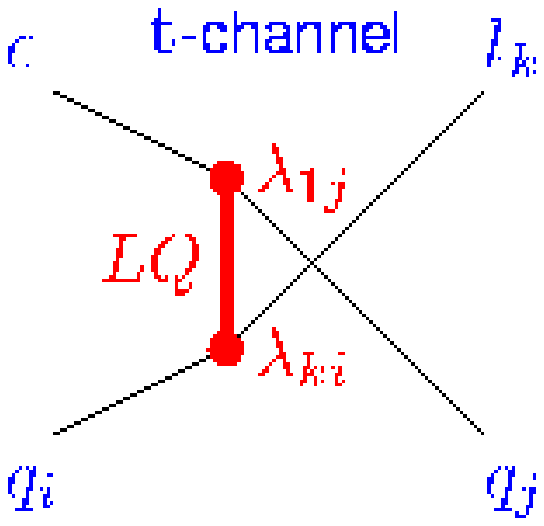,height=3.cm}
$$
\vspace{-1.cm}
\caption{ \it LQ production (exchange) in the $s$ ($t$) channel.}
\label{fig:lqdiag}
\end{figure}

In various models (e.g. some models of Grand Unification),
triplets of coloured bosons, called leptoquarks (LQs),
are supposed to be produced, which could be observed at HERA by an $eq$ fusion as
illustrated in figure \ref{fig:lqdiag}.

LQs are scalar or vector bosons 
characterized by a weak Isospin $T=0,~1/2,~1$ and
a fermionic number $F=0,~2$ (defined as $F=L+3B$ where $L$ and
$B$ are respectively the lepton and baryon numbers).
The model proposed by Buchm\"uller, R\"uckl and Wyler (BRW)
\cite{brw} describes 14 LQs
(each coupling to only one combination of fermion chiralities):

\begin{itemize}
\item 7 LQs with a fermion number $F=0$ searched for in $e^+p$
collisions.
\item 7 LQs with a fermion number $|F|=2$ searched for in $e^-p$ data.
\end{itemize}

In the BRW model, first generation LQs decay into $eq$ and $\nu q$ with fixed
branching ratios.

As the LQs decay topologies are identical to the Deep Inelastic Scattering (DIS) 
processes, LQs 
decaying into $eq$ are searched for in NC samples
whereas LQs decaying into $\nu q$ are searched for in Charged
Current (CC) samples.

\begin{figure}[htbp] 
$$
\epsfig{figure=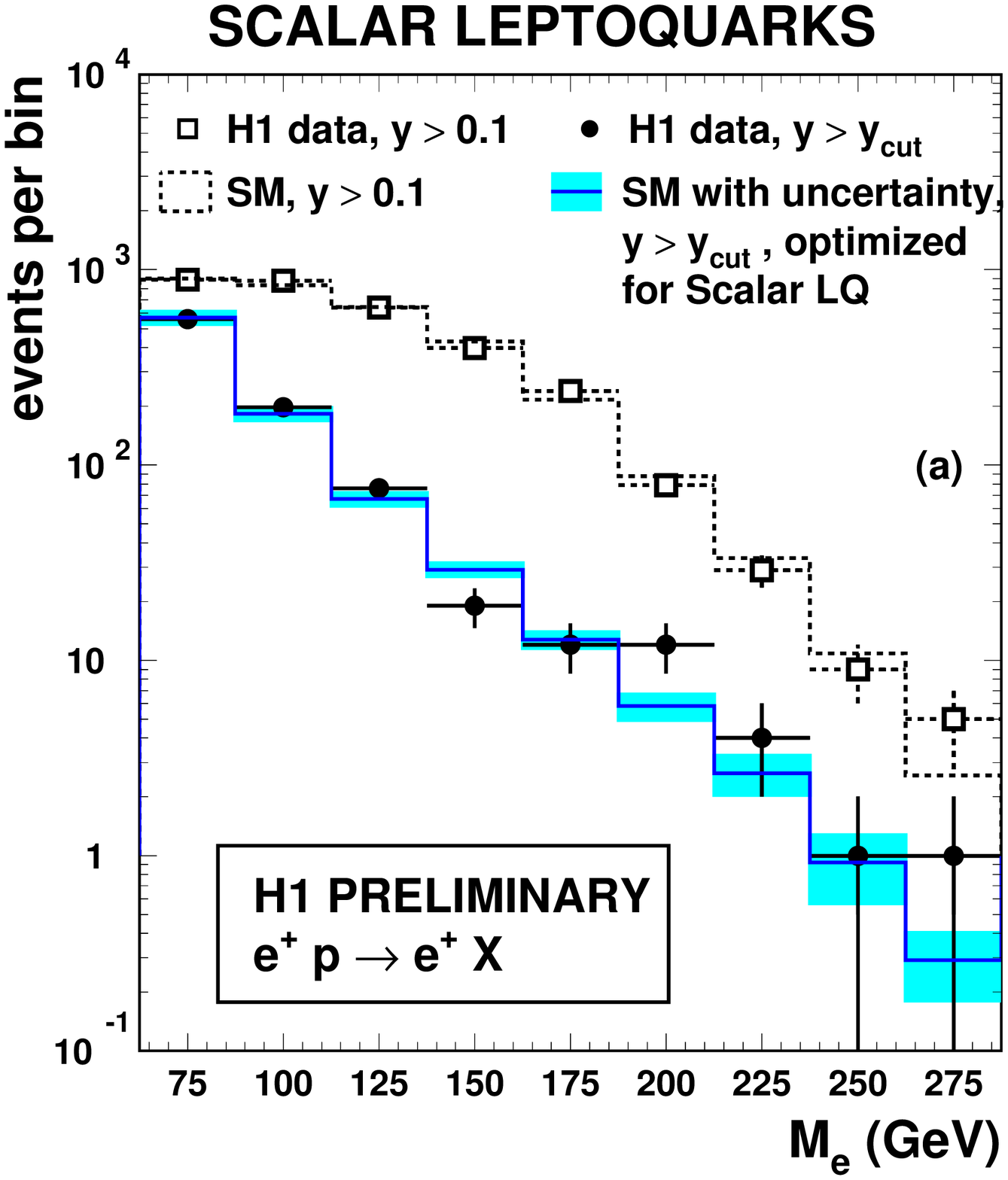,height=6.cm}
\epsfig{figure=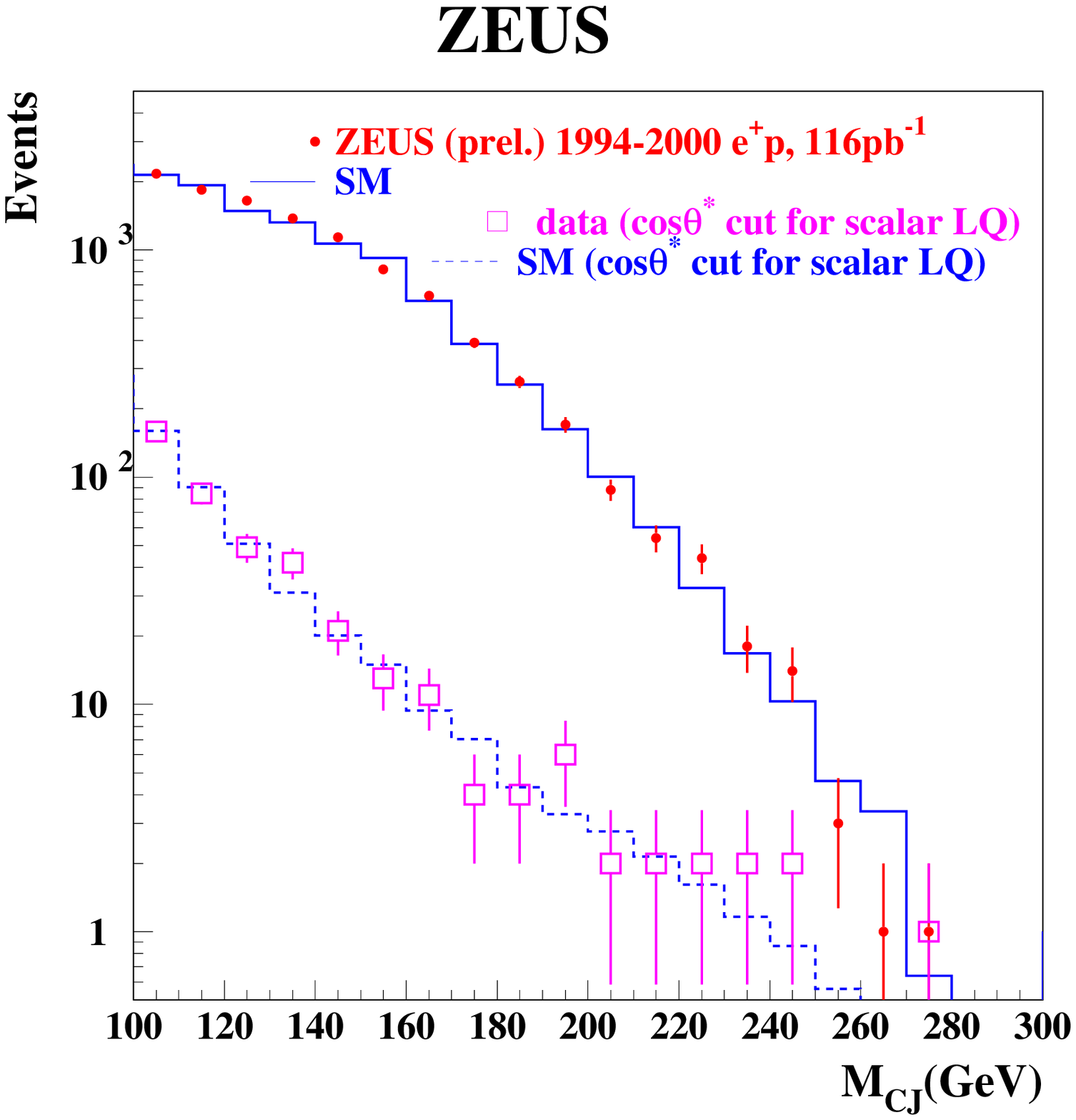,height=6.cm}
$$
\vspace{-1.2cm}
\caption{ \it Mass distributions in the NC channel for H1 (left) 
and ZEUS (right) experiments with 94-00 $e^+p$ data, with and without the
angular cut.}
\label{fig:lqnc}
\end{figure}

\begin{figure}[htbp] 
\vspace{-0.3cm}
$$
\hspace{-7.cm}
\epsfig{figure=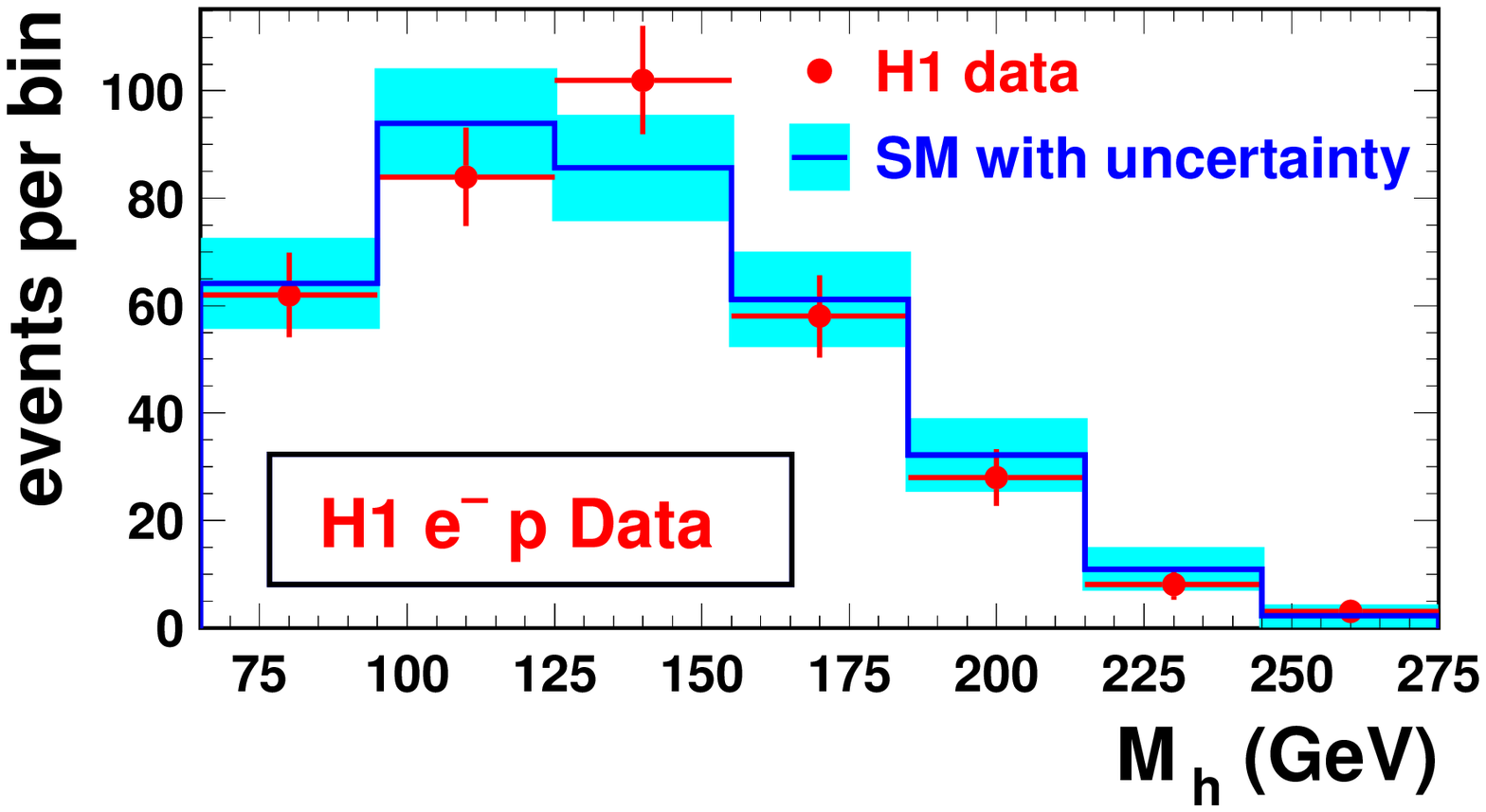,height=10.cm}
$$
\vspace{-11.cm}
$$
\hspace{10.cm}
\epsfig{figure=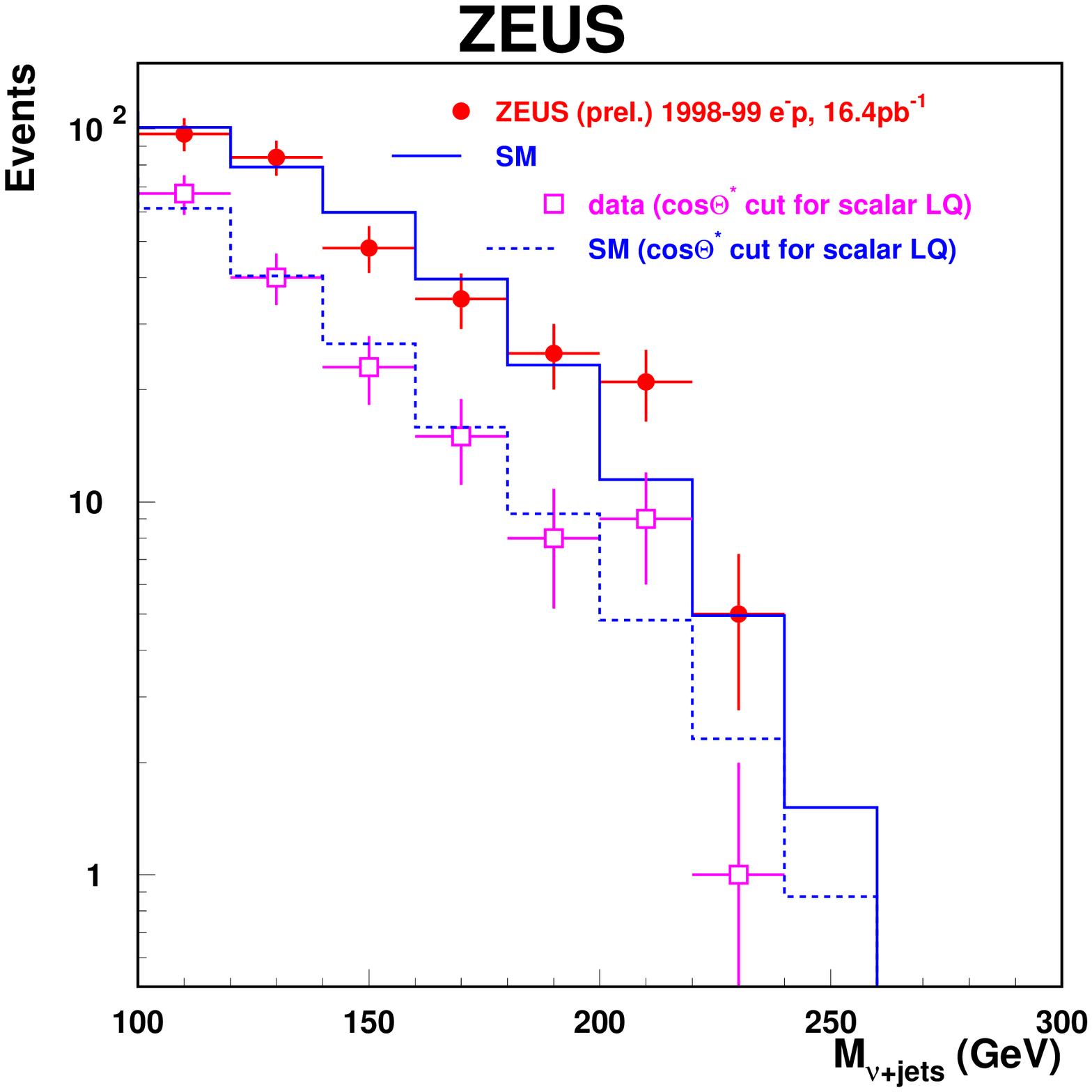,height=6.cm}
$$
\vspace{-1.4cm}
\caption{ \it Mass distributions in the CC channel for H1 (left) 
and ZEUS (right) experiments with 98-99 $e^-p$ data. ZEUS applies an additional
angular cut as a function of the LQ mass.}
\label{fig:lqcc}
\end{figure}

The presence of LQs would manifest itself as an excess of DIS-like events with
a large angle for the final state lepton.
Therefore an angular cut varying with the LQ mass is applied in order to
optimize the sensitivity to LQs.
Mass distributions in the NC and CC channels are respectively shown in figures
\ref{fig:lqnc} and \ref{fig:lqcc} \cite{h1lq,h1lqpap,zeuslq}.

In the NC channel, the previous excess observed 
at masses around 200 GeV in 
H1 and ZEUS 1994-97 data has not been confirmed by new 1999-2000 data. Therefore
the mass distribution in the full data sample
is compatible with the SM expectation.
In the CC channel, both experiments show also a good agreement with the SM
expectation.
Therefore constraints on LQ masses and their coupling $\lambda$ to $eq$ pairs
are derived.

\begin{figure}[htbp] 
\vspace{-0.5cm}
$$
\epsfig{figure=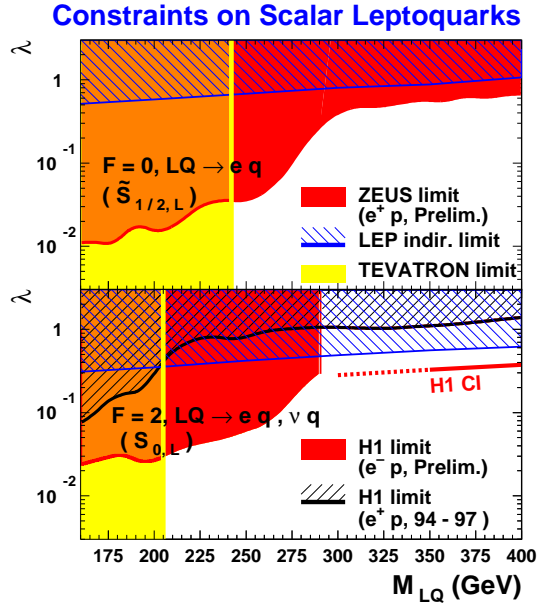,height=8.6cm}
$$
\vspace{-1.2cm}
\caption{ \it Limits on the coupling $\lambda$ as a function of the LQ mass for
fermion numbers $F=0$ and $F=2$ for scalar LQs, with HERA, LEP and Tevatron
colliders.}
\label{fig:lqliml}
\end{figure}

Figure \ref{fig:lqliml} presents the resulting limits on the coupling
$\lambda$ as a function of the LQ mass for scalar LQs, in the context of 
the BRW model (fixed branching ratios). HERA limits are seen to be competitive
with those obtained at other colliders.
Moreover HERA extends the exclusion region at high mass of the LQ.

In the more general case where the branching ratio $\beta_{eq}$ ($\beta_{\nu q}$)
for the LQ decay into $eq$ ($\nu q$) is not fixed anymore, constraints on 
$\beta_{eq}$ can be set as a function
of the LQ mass, for different values of the coupling $\lambda$.

\begin{figure}[htbp] 
$$
\hspace{-0.7cm}
\epsfig{figure=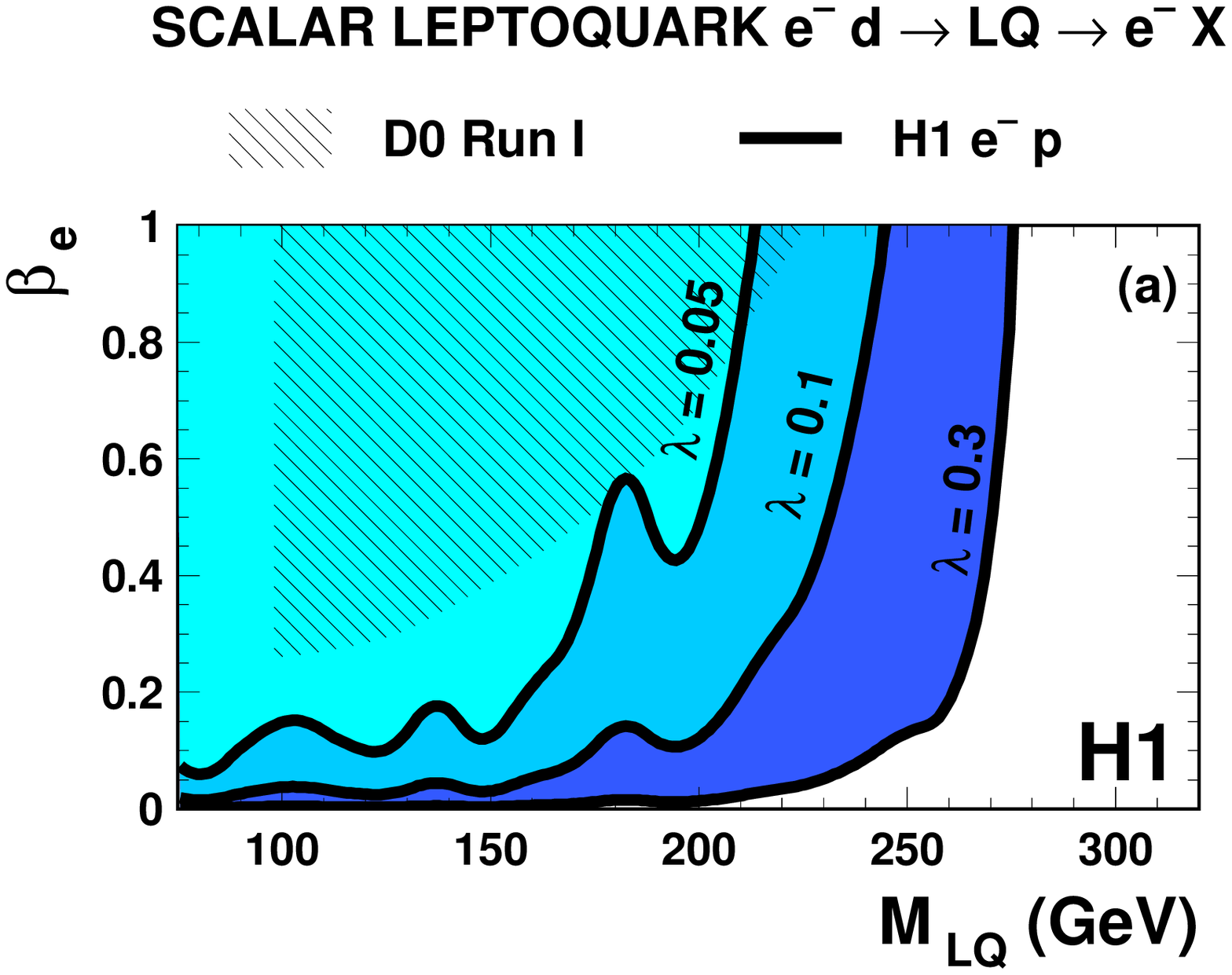,height=6.cm}
\epsfig{figure=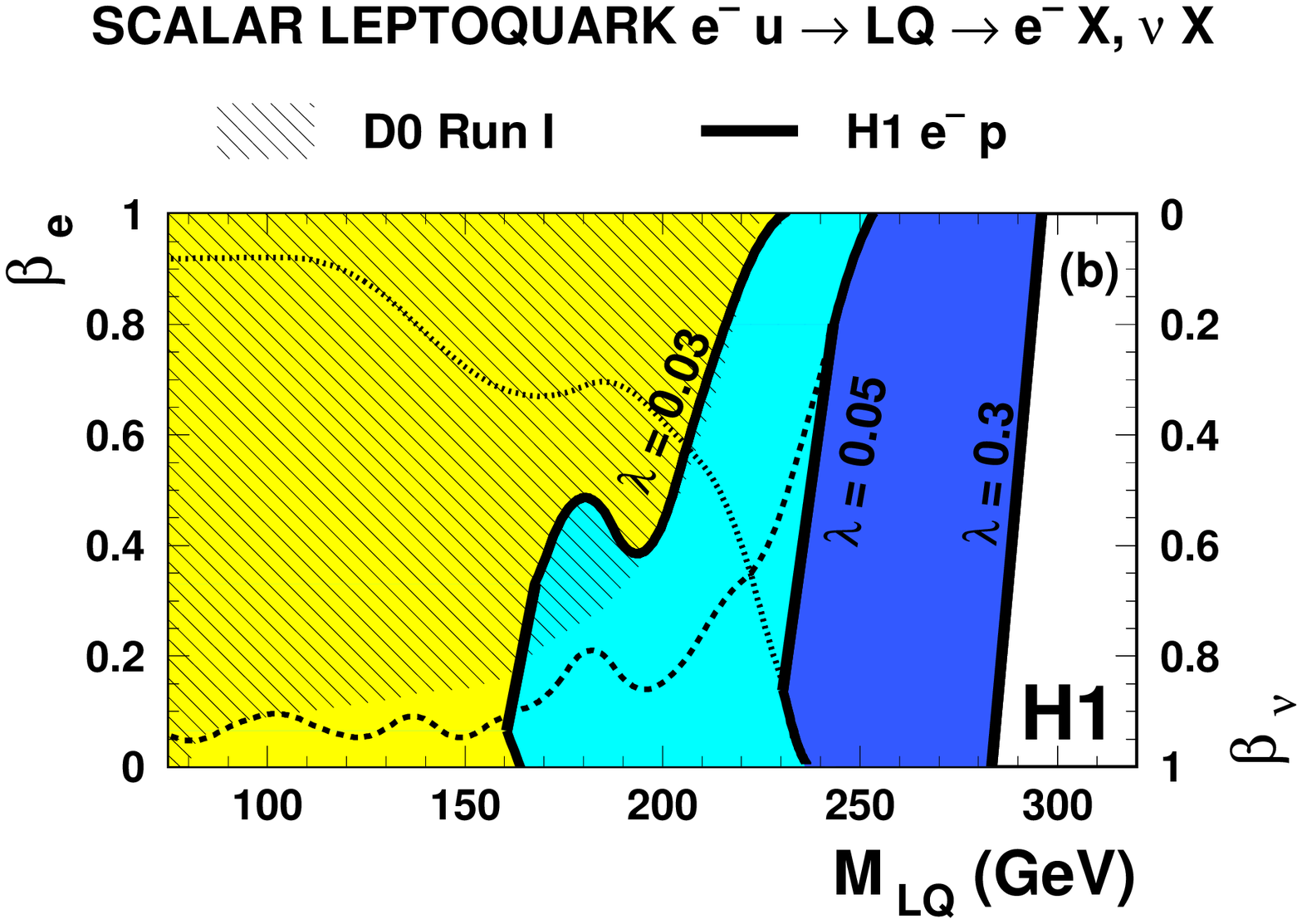,height=6.cm}
$$
\vspace{-1.3cm}
\caption{ \it The left (right) plot shows limits on the branching ratio 
$\beta_{e}$ versus the LQ
mass for scalar LQs in the NC (NC+CC) channel
with H1 data. The hashed region 
is the region excluded by Tevatron.
The domains to the left of the solid curves are excluded by H1. In the second
figure, the limits combine the NC and CC channels.
For $\lambda=0.5$, the domain excluded by the NC (CC) channel alone is also
shown above (below) the dashed (dotted) curve.}
\label{fig:other}
\end{figure}

H1 results are shown in figure 
\ref{fig:other} for two example scalar LQs.
This shows that HERA limits are compatible with Tevatron for
low values of $\lambda$ and superior at large $\lambda$. Moreover HERA extends
the limits to small values of $\beta_{eq}$.

\section{Lepton-Flavour Violation}

Experiments searching for neutrino oscillations show an increasing evidence
for Lepton Flavour Violation, therefore the lepton number is 
likely to be violated.
Processes like $eq\rightarrow \mu q'$ or
$eq\rightarrow \tau q'$ via a LQ production or exchange are searched for
(these processes are also illustrated by figure \ref{fig:lqdiag}).
Both these transitions have been looked for in various data sets,
and constraints have been set \cite{h1lfvpast,zeuslfvpast}.

As an example, constraints on the process $eq \rightarrow \mu q'$ which involves
two couplings $\lambda_{eq}$ and $\lambda_{\mu q}$ are illustrated here.

\begin{figure}[htbp] 
$$
\epsfig{figure=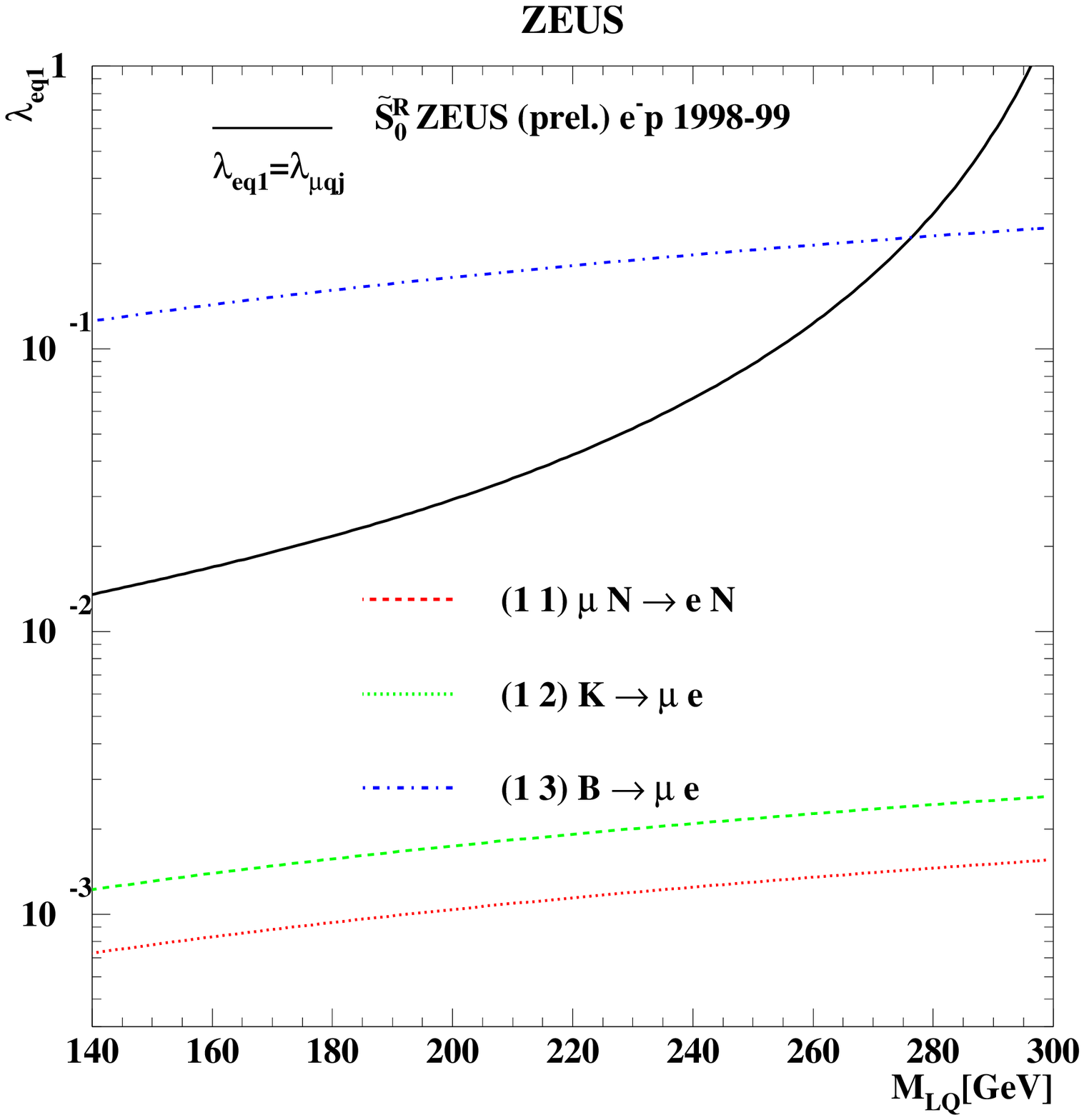,height=7.5cm}
\epsfig{figure=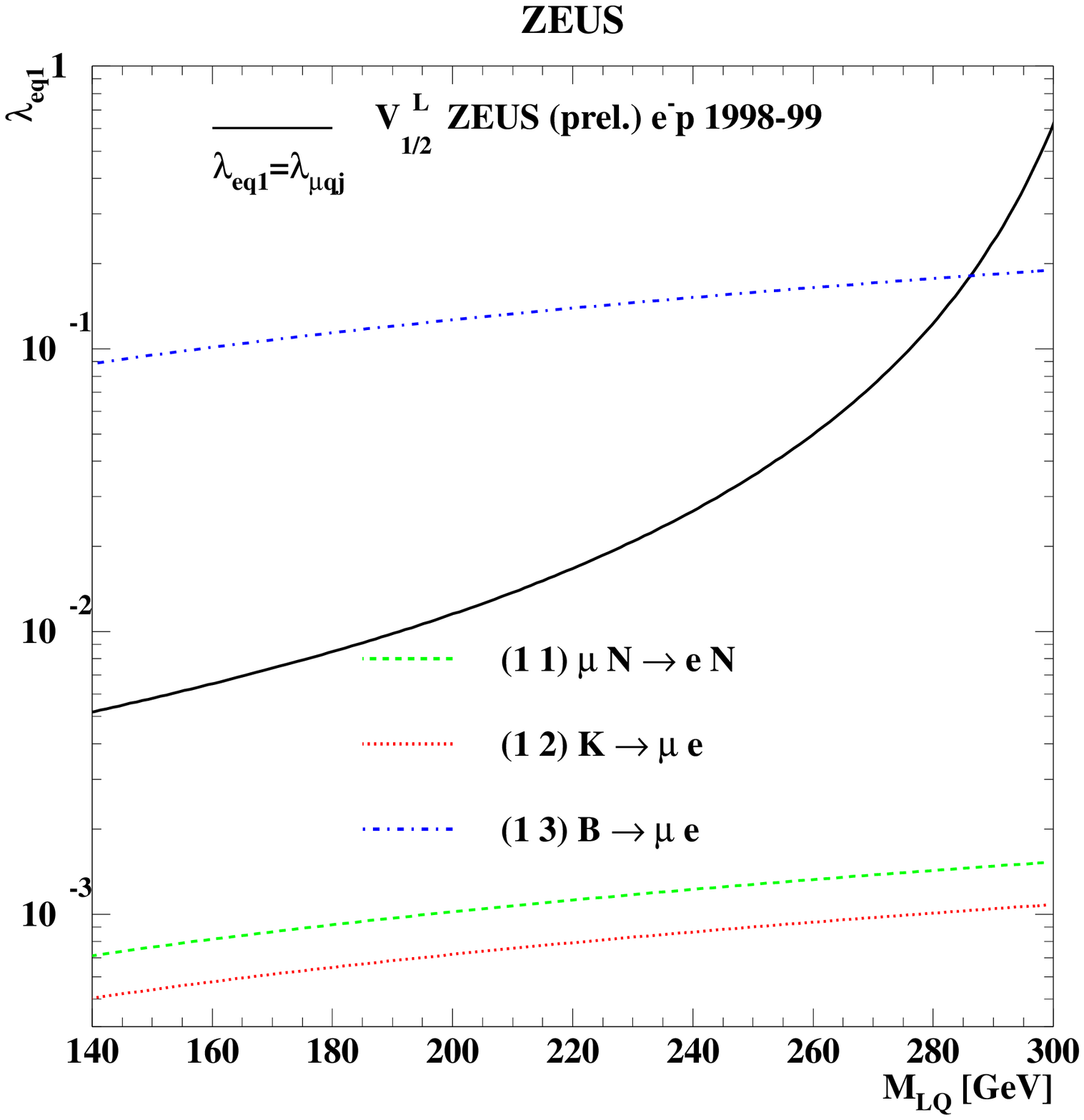,height=7.5cm}
$$
\vspace{-1.cm}
\caption{ \it The left (right) plot shows limits on the coupling constant 
$\lambda_{eq}$ as a function of the LQ mass for a scalar (vector) LQ
exchange from the $e^-p$ 1998-99 ZEUS data.}
\label{fig:lfv}
\end{figure}

At low LQ mass, upper limits on the product 
$\lambda_{eq} \times \sqrt{BR_{LQ\to \mu q}}$ can be deduced as a function
of the LQ mass.
Assuming $\lambda_{eq}=\lambda_{\mu q}$, the branching ratio 
$BR_{LQ\to \mu q}$ is fixed to 0.5 and limits on $\lambda_{eq}$ 
versus the LQ mass are shown in figure \ref{fig:lfv} \cite{zeuslfv}.
This figure shows that HERA limits are better than low-energy experiments
for $ed\rightarrow \mu b$ transitions via the process $B \rightarrow \mu e$
and for a LQ mass lower than 280 GeV.

At high LQ mass, the cross section is proportional
to $\left( \frac{\lambda_{eq} \lambda_{\mu q}}{M^2} \right)^2$ where $M$ is
the LQ mass. Bounds on this ratios were also set in 
\cite{h1lfvpast} and \cite{zeuslfvpast}.

\section{Squarks produced by R-parity violation}

In the Minimal Supersymmetric Standard Model (MSSM) \cite{mssm},
R-parity is defined
as $R_p=(-1)^{3B+L+2S}$. SM particles have $R_p=1$
whereas their supersymmetric partners have $R_p=-1$.

\begin{figure}[hhhh] 
$$
\epsfig{figure=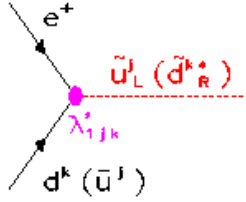,height=3.5cm}
$$
\vspace{-1.cm}
\caption{ \it Squark production at HERA. $\lambda'$ is the \rp
coupling.}
\label{fig:sqgdiag}
\end{figure}

If $R_p$-violation would be allowed,
superparticles could be produced at HERA by an $eq$ fusion: 
$ep \rightarrow \tilde{q}X$. This process is illustrated by figure
\ref{fig:sqgdiag}. An $R_p$-violating coupling $\lambda'$ is involved
corresponding to an
$R_p$-violation term $\lambda'_{ijk} L_i Q_j \overline{D}_k$
added to the lagrangian.

The $e^+p$ data can mainly probe the coupling $\lambda'_{1j1}$
via the process $e^++d \rightarrow \tilde{u}_L$.

Both decay modes of the squark are investigated at HERA: \rp decays and
``gauge'' decays.

\begin{figure}[htbp] 
$$
\hspace{-6.cm}
\epsfig{figure=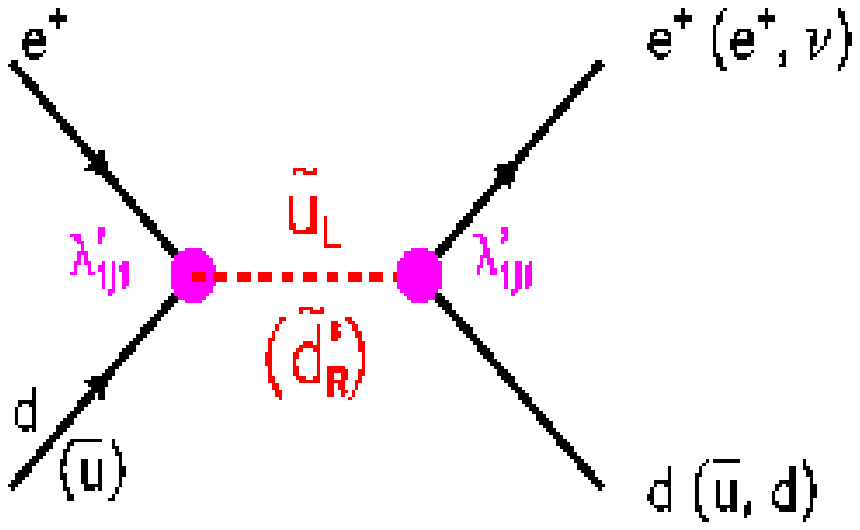,height=3.5cm}
$$
\vspace{-1.cm}
\caption{ \it Squark decay by $R_p$-violation
and expected signature and main backgrounds from SM processes.}
\label{fig:rpdiag}
\vspace{-4.cm}
\hspace{9.cm}
\begin{tabular}{|c|c|}
\hline
Signature & Main Bkg \\
\hline
$e+$jet & NC \\
\hline
$\nu+$jet & CC \\
\hline
\end{tabular}
\end{figure}

The \rp decay mode is illustrated in figure \ref{fig:rpdiag}.
As the squark can decay into an electron and a quark or a neutrino
and a quark, the main backgrounds are respectively NC and CC processes.

\vspace{3.cm}

\begin{figure}[htbp] 
$$
\hspace{-7.cm}
\epsfig{figure=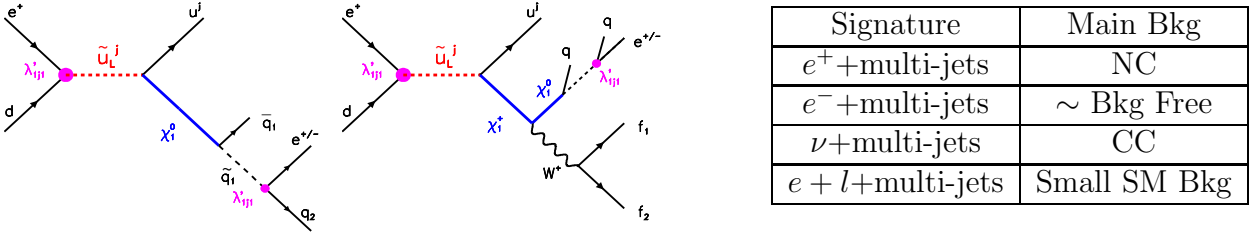,height=3.5cm}
$$
\vspace{-1.cm}
\caption{ \it ``Gauge'' decay of the squark into a neutralino (left) or
chargino (right). When the final lepton 
(arising from the $R_p$-violating neutralino decay)
and incident beam lepton have different signs, 
the channel is almost background-free.}
\label{fig:gbdiag}
\vspace{-5.2cm}
\hspace{10.cm}
\begin{tabular}{|c|c|}
\hline
Signature & Main Bkg \\
\hline
$e^++$multi-jets & NC \\
\hline
$e^-+$multi-jets & $\sim$ Bkg Free \\
\hline
$\nu+$multi-jets & CC \\
\hline
$e+l+$multi-jets & Small SM Bkg \\
\hline
\end{tabular}
\end{figure}

Example final states resulting from a squark ``gauge'' decay (mediated
by a $R_p$-conserving coupling) are illustrated in figure  \ref{fig:gbdiag}.
The table indicates all signatures considered, together with their
corresponding backgrounds for $e^+p$ scattering.
No deviation from the SM has been observed, therefore limits
on the coupling $\lambda'$ are derived.

\vspace{4.2cm}

{\tiny .}
\vspace{2.cm}

\begin{figure}[hhhh] 
\vspace{-1.cm}
$$
\hspace{-3.5cm}
\epsfig{figure=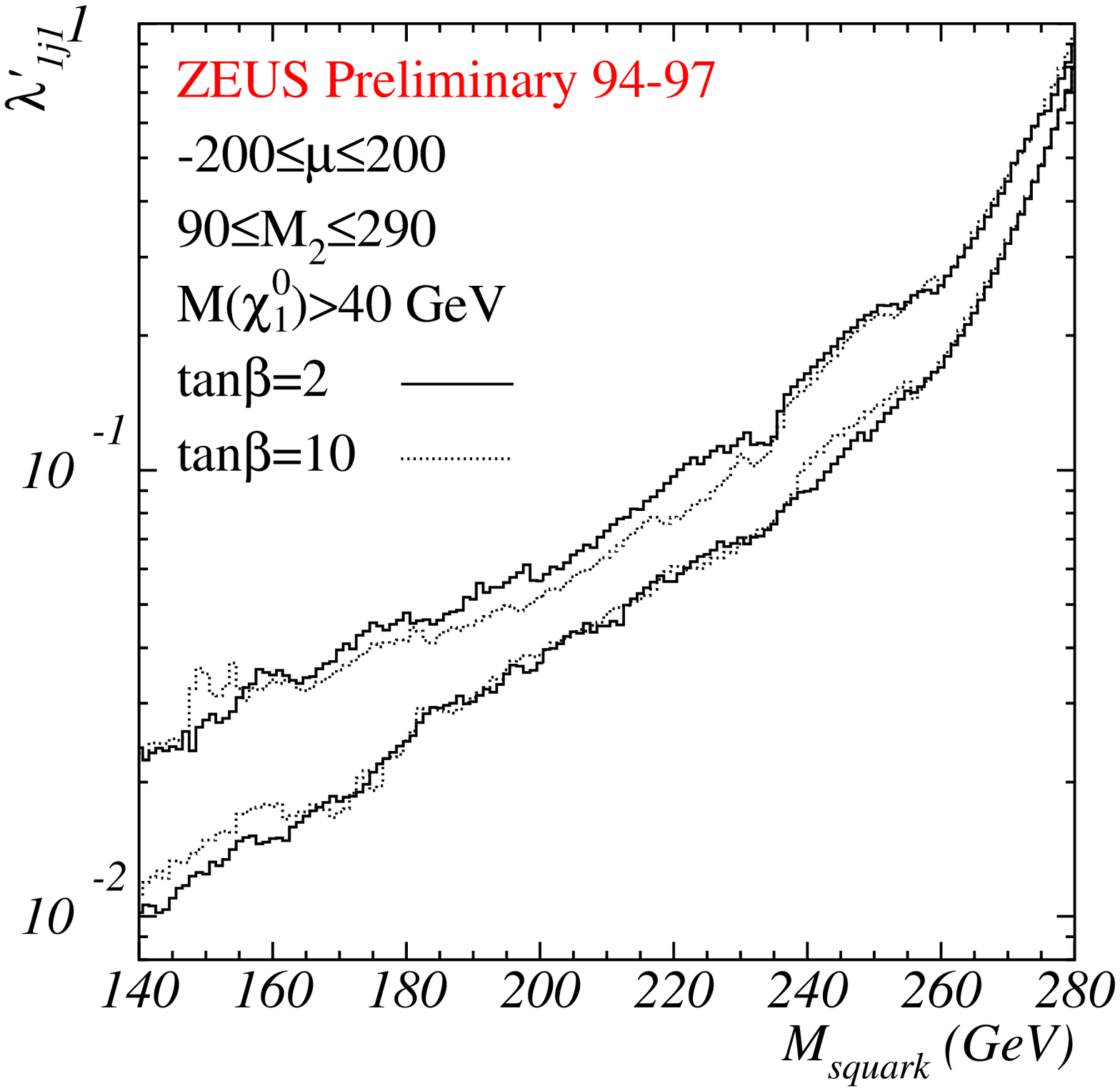,height=5.5cm}
$$
\vspace{-5.7cm}
$$
\hspace{7.5cm}
\epsfig{figure=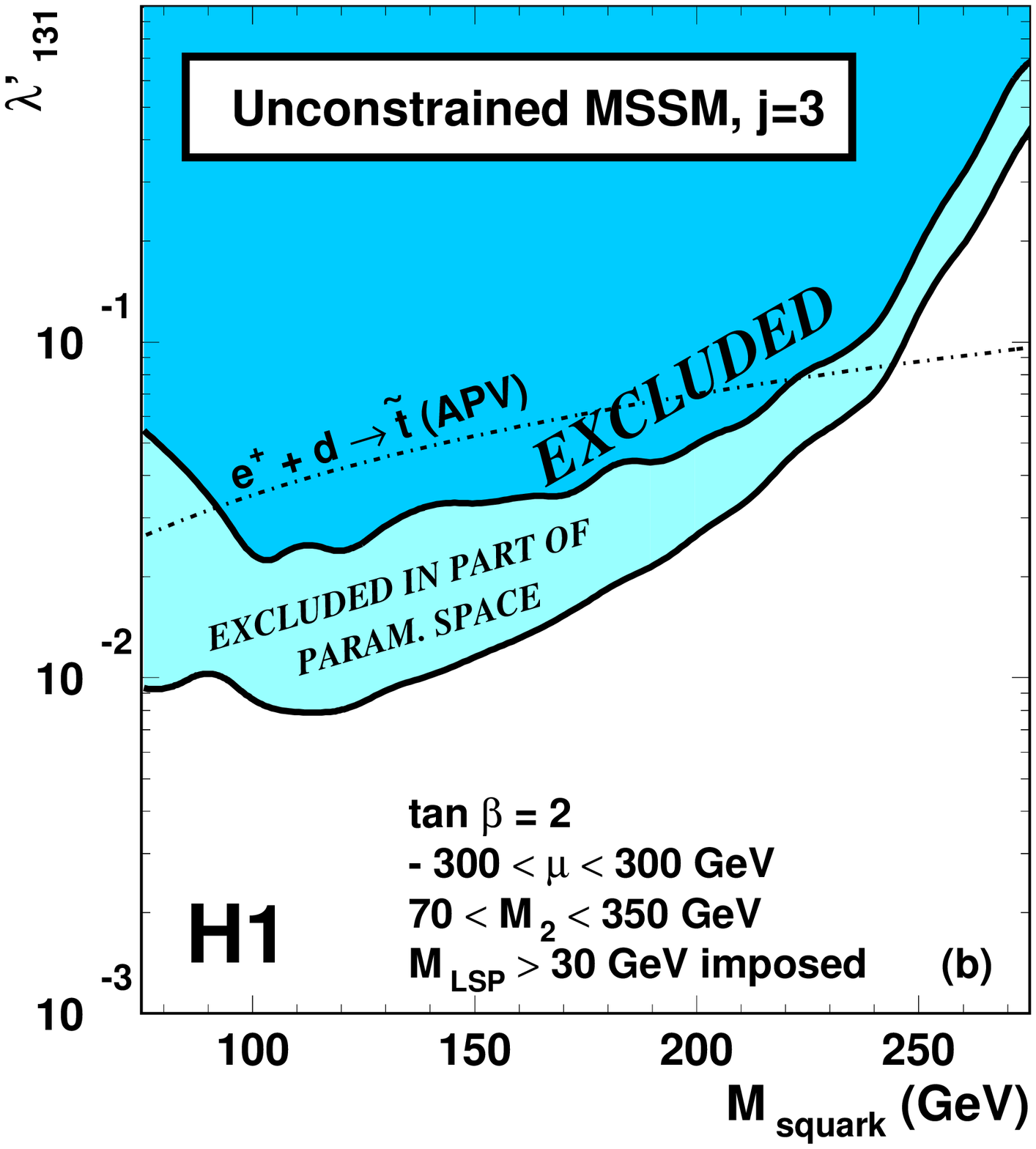,height=6.cm}
$$
\vspace{-0.5cm}
\caption{ \it Left (right) plot shows limits on $\lambda'_{111}$
and $\lambda'_{121}$ ($\lambda'_{131}$) obtained by the ZEUS (H1) 
experiment versus the squark mass.
The two curves represent for a given value of $\tan{\beta}$ the minimum
and maximum $95\%$ CL limits obtained when scanning on $M_2$ and $\mu$.
In the right plot the colored regions represent the
H1 exclusion region while the dashed curve shows the limits from
Atomic Parity Violation.}
\label{fig:susylim1}
\end{figure}

The results have been interpreted in various SUSY models.
The first one is the unconstrained MSSM.
In this model, sfermion masses are considered as free parameters.
The other free parameters are: the squark mass $M_{\tilde{q}}$,
the soft SUSY breaking mass term $M_2$, the mixing mass term for 
Higgs doublets  $\mu$ and the ratio of the 
vacuum expectation values of the 2 neutral Higgs bosons $\tan{\beta}$.
A scan of the SUSY parameter space $(M_2,\mu,\tan{\beta})$ is performed.
The resulting limits on $\lambda'$ of the H1 and ZEUS experiments are shown in
figure \ref{fig:susylim1} \cite{h1squarkpap,zeussquarkpap}.

As they are calculated for a large range
in $\mu$ and $M_2$, these limits are almost model independent. They are seen
to be more restrictive than the indirect bounds from Atomic Parity Violation
(APV) for the coupling $\lambda'_{131}$, which would lead to stop production
at HERA.
For a given value of 
$\lambda_{1j1}'=0.3$, squark masses below 260 GeV are excluded.

Another SUSY model is the minimal Super-Gravity model
(mSUGRA) \cite{msugra} which contains the additional
constraint of a universal mass $m_0$ introduced for all sfermions.
In this model the electroweak symmetry breaking is
supposed to be driven by radiative corrections. Therefore
the free parameters are mainly $m_0$ and the unified gaugino mass
$m_{1/2}$ at the GUT scale, the sign of $\mu$ and $\tan{\beta}$. 

\begin{figure}[hhhh] 
$$
\epsfig{figure=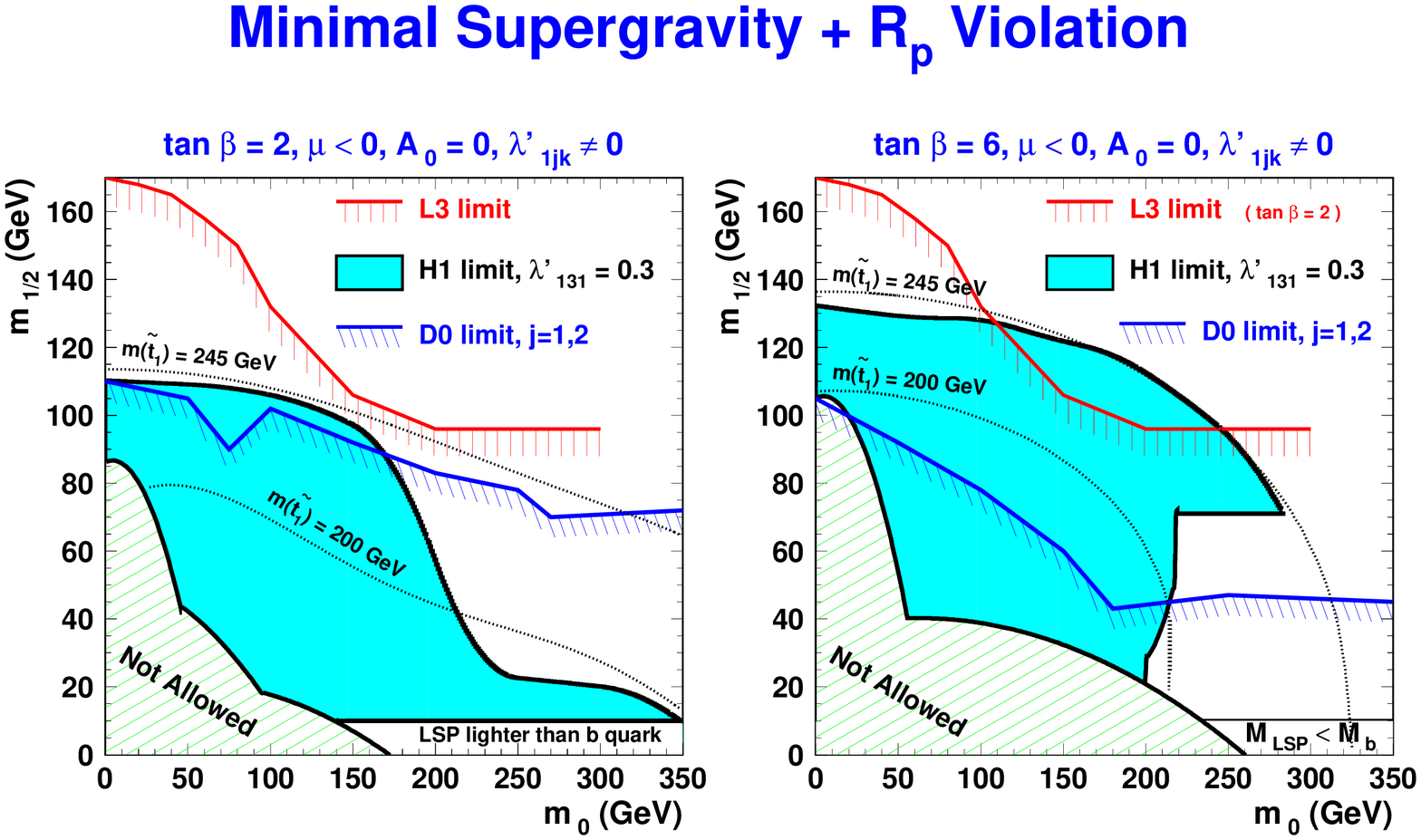,height=8.cm}
$$
\vspace{-1.cm}
\caption{ \it H1 exclusion regions for a coupling  $\lambda'_{131}=0.3$ for 
$\tan{\beta}=2$ (left) and $\tan{\beta}=6$ (right) in the plane ($m_0,m_{1/2}$).
$\lambda'_{1jk}$ obtained by D0 and L3 are also shown.}
\label{fig:msugra}
\end{figure}

The resulting limits in the plane ($m_0,m_{1/2}$)
 are presented in figure
\ref{fig:msugra}. H1 limits are competitive with (better than) Tevatron for 
low (medium) $\tan{\beta}$.
For the coupling $\lambda_{131}'$, H1 bounds are comparable to LEP limits\footnote{
LEP limits should not depend too much on $\tan{\beta}$ for small and medium
values of $\tan{\beta}$.}
for intermediate values of $m_0$, $m_{1/2}$ and $\tan{\beta}$.

\section{Excited fermions}

\begin{figure}[htbp]
$$
\epsfig{figure=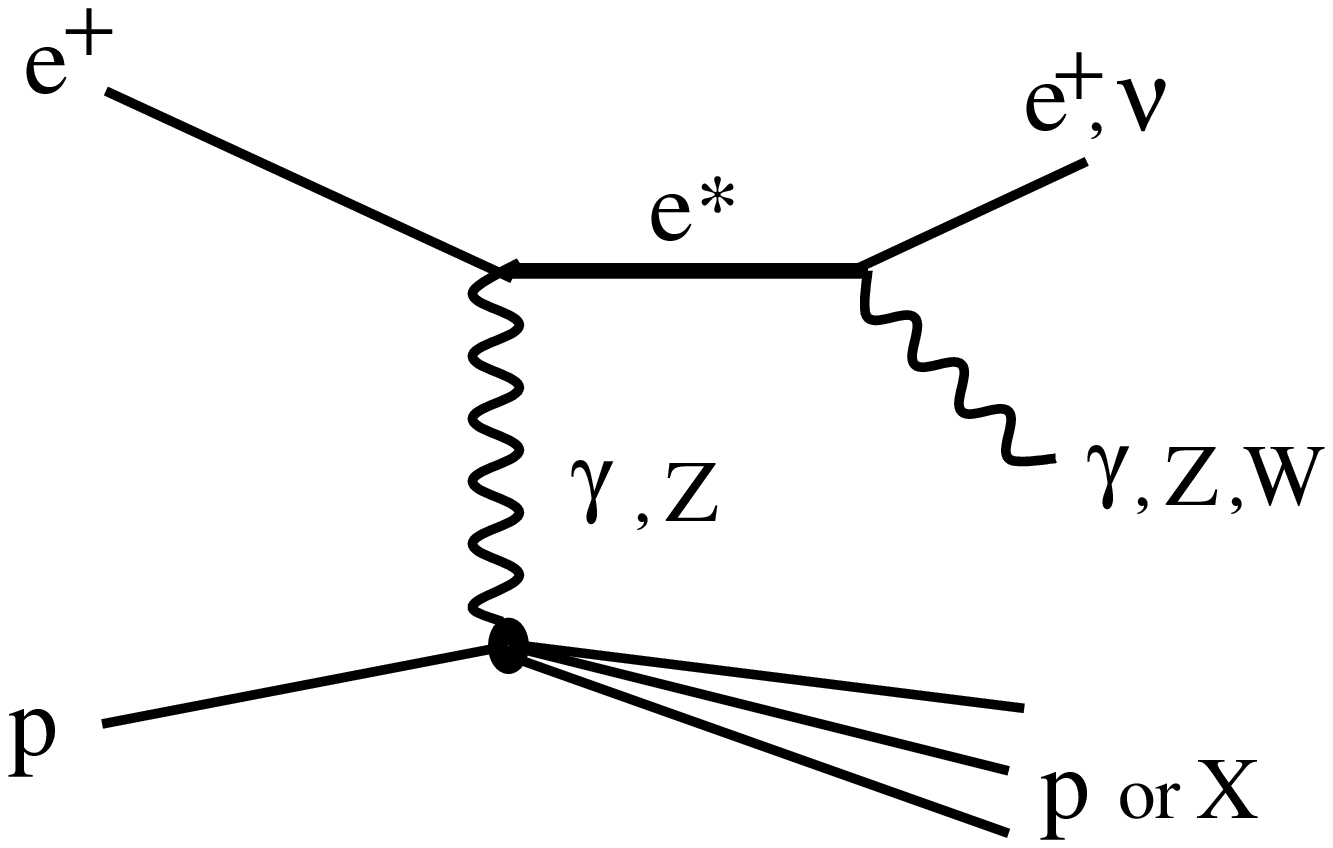,height=3.2cm}
\epsfig{figure=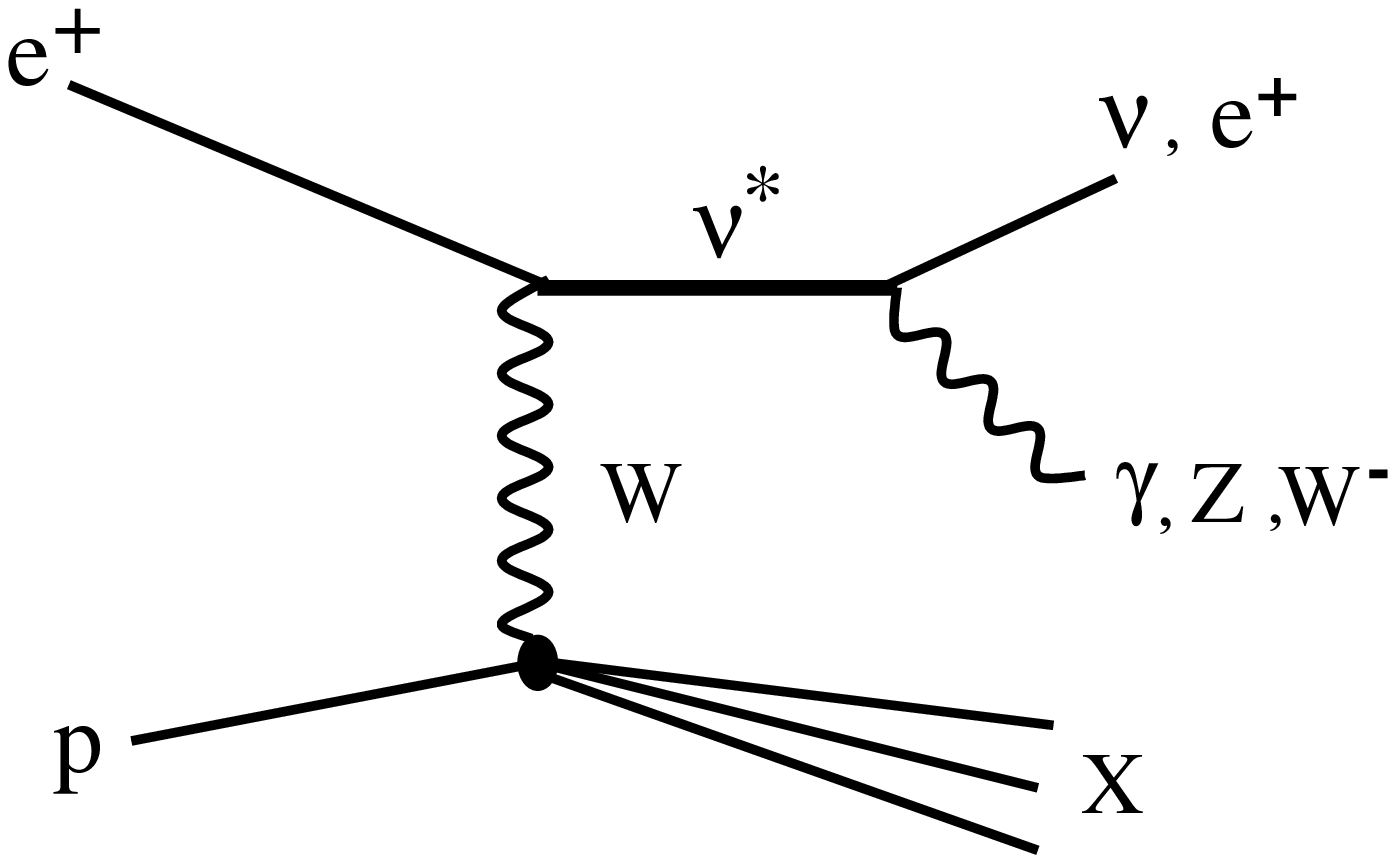,height=3.2cm}
\epsfig{figure=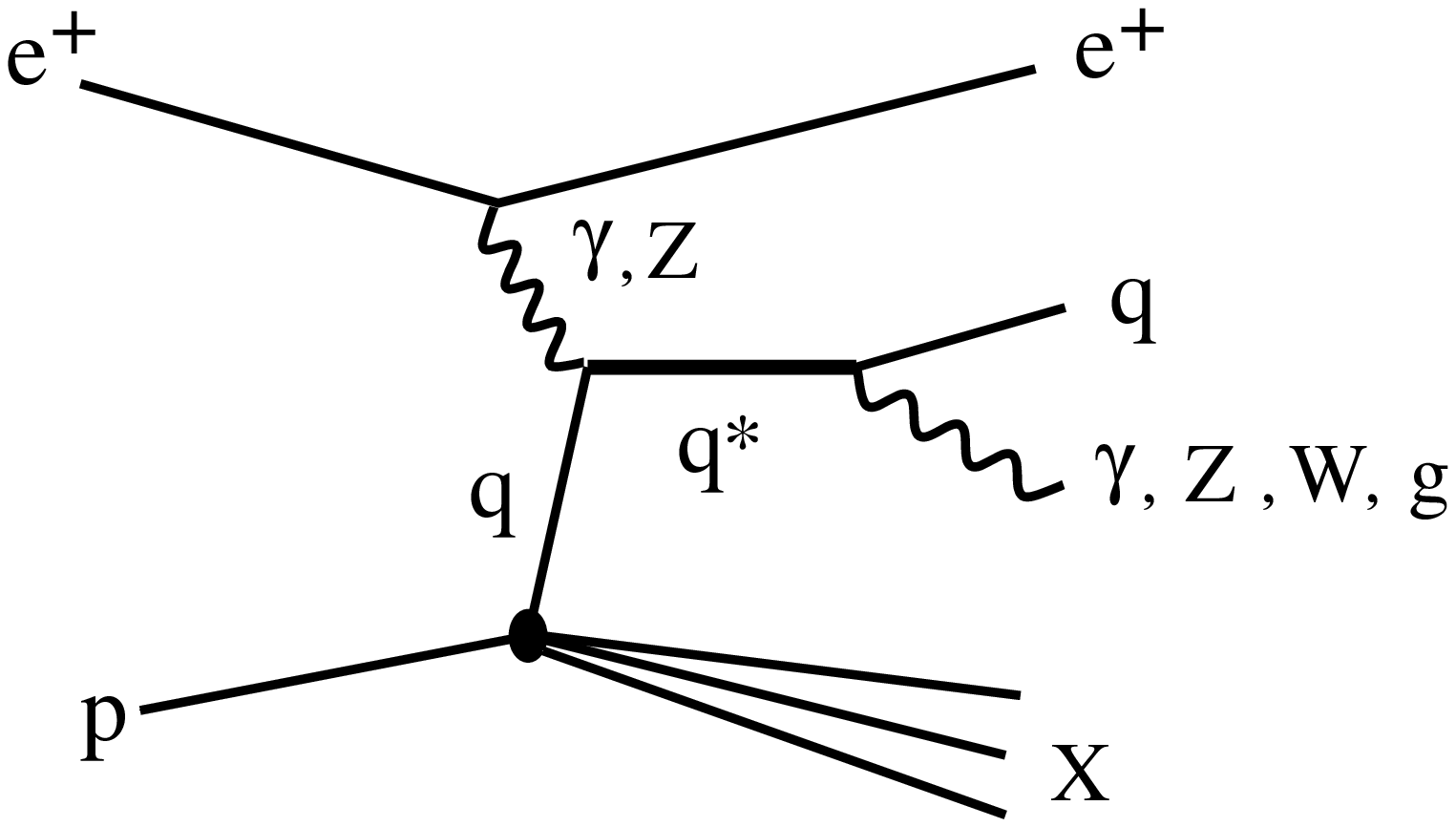,height=3.2cm}
$$
\vspace{-1.cm}
\caption{ \it Excited fermion production at HERA}
\label{fig:fdiag}
\end{figure}

In various extensions of SM, excited states of fermions are expected
($e^*$, $\nu^*$ and $q^*$).
The excited fermions can then de-excite by emission
of electroweak bosons $\gamma$, $Z$ or $W$ as shown in figure 
\ref{fig:fdiag} producing various final state topologies to be searched for.
This process is described
by the effective lagrangian (\ref{eq:felag}) \cite{excited}.

\begin{equation} \label{eq:felag}
{\cal L}_{int}=\frac{1}{2\Lambda}
\overline{F}_R^*\sigma^{\mu \nu} \left[
g f \frac{\tau^a}{2} W_{\mu \nu}^a
+ g' f'\frac{Y}{2} B_{\mu \nu}
+ g_s f_s\frac{\lambda^a}{2} G_{\mu \nu}^a
\right] F_L
\end{equation}

\vspace{0.5cm}
$F_R^*$ and $F_L^*$ are
the left-handed and right-handed components of the excited form weak
isodoublets. $W_{\mu \nu}^a$, $B_{\mu \nu}$ and
 $G_{\mu \nu}^a$ are the field-strength tensors of the respective gauge groups 
SU(2), U(1) and SU(3). $g$, $g'$ and $g_s$ are the SM
couplings. $f$, $f'$ and $f_s$ determine the coupling strength between the
excited fermions and the bosons corresponding to the gauge groups 
U(1), SU(2) and SU(3).
$\tau^a$ are the Pauli matrices, $Y$ the weak hypercharge operator,
$\lambda^a$ the Gell-Mann matrices and $\Lambda$ the compositeness scale.

Assuming relations between $f$, $f'$ and $f_s$, the branching ratios are fixed
and limits on the ratio $f/\Lambda$ versus the excited fermion mass can be
deduced from measurements.

\begin{figure}[htbp]
$$
\hspace{-0.5cm}
\epsfig{figure=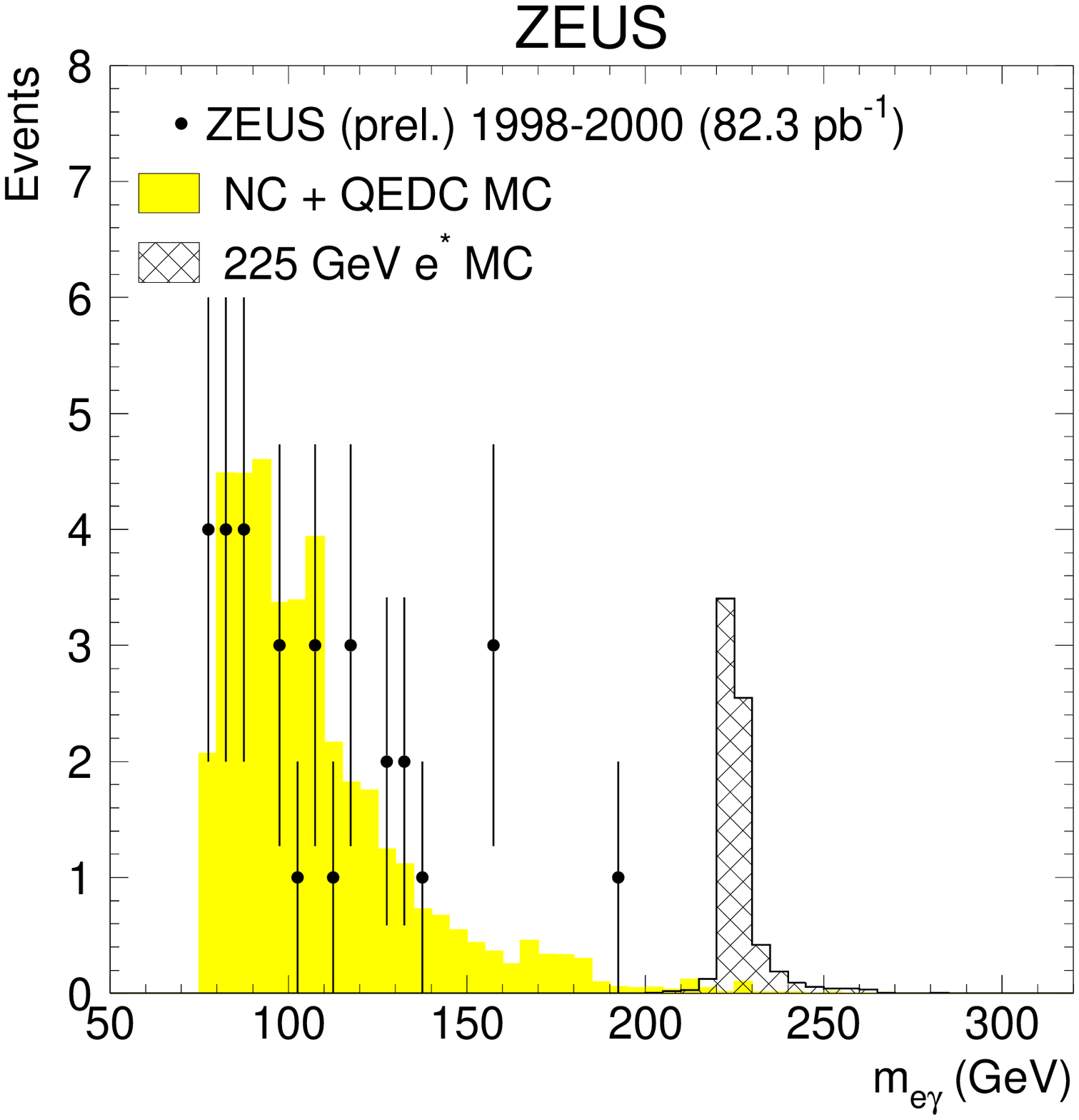,height=5.5cm}
\hspace{-0.5cm}
\epsfig{figure=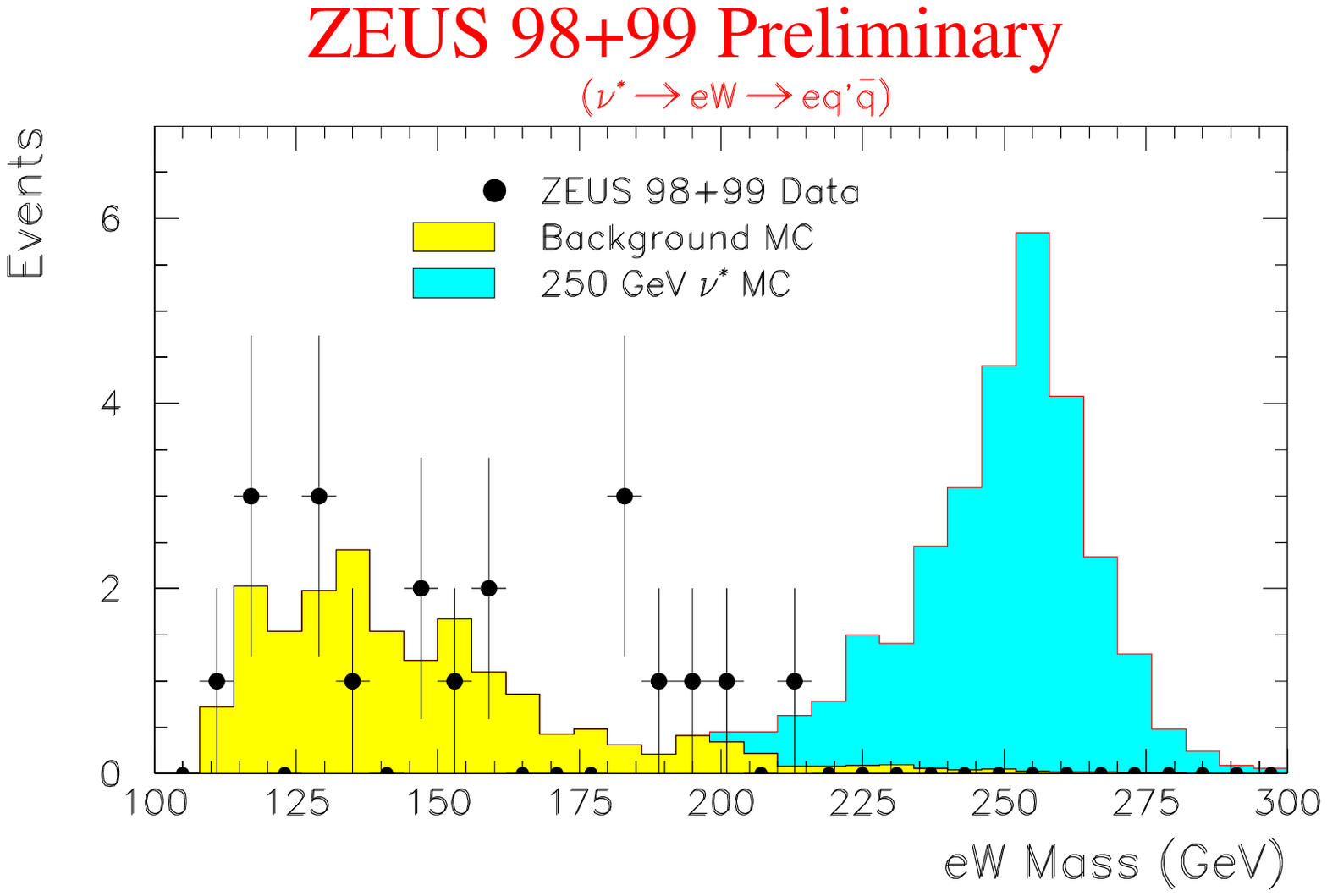,height=5.cm}
\hspace{-0.5cm}
\epsfig{figure=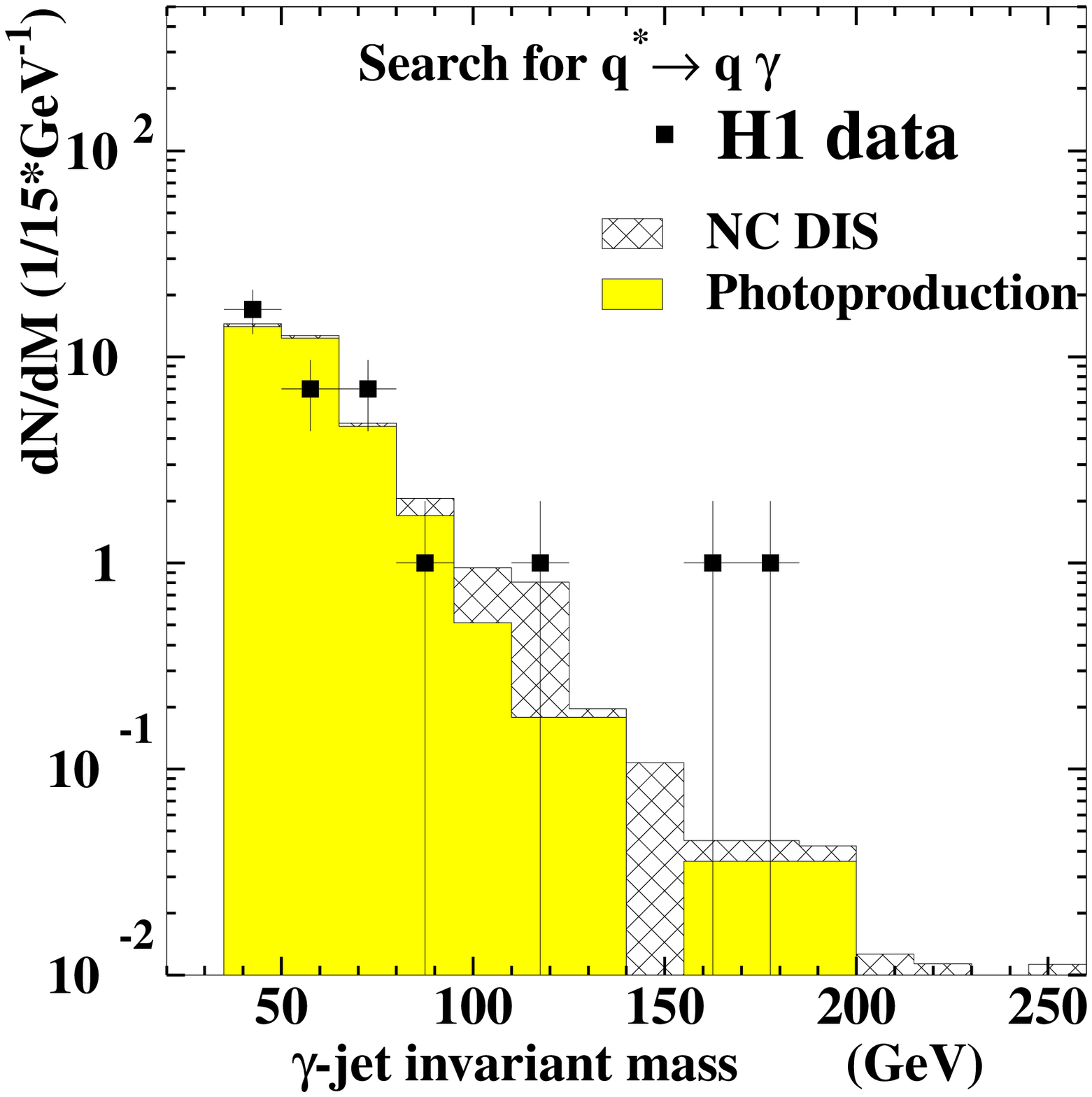,height=5.5cm}
$$
\vspace{-1.cm}
\caption{ \it Mass distribution of the $e^*$, $\nu^*$ and
$q^*$ candidates. Data samples are respectively 98-00 $e^\pm p$ (ZEUS),
98-99 $e^-p$ (ZEUS) and 94-97 $e^+p$ (H1).}
\label{fig:efmass}
\end{figure}

Figure \ref{fig:efmass} shows the mass distribution of the excited
electron, neutrino and quark candidates in $e \gamma$, $eW$ and $q\gamma$ channels.
\cite{h1excitedpap,zeusexcited}. No deviation from the SM
is observed, therefore limits are established.

\begin{figure}[htbp]
$$
\epsfig{figure=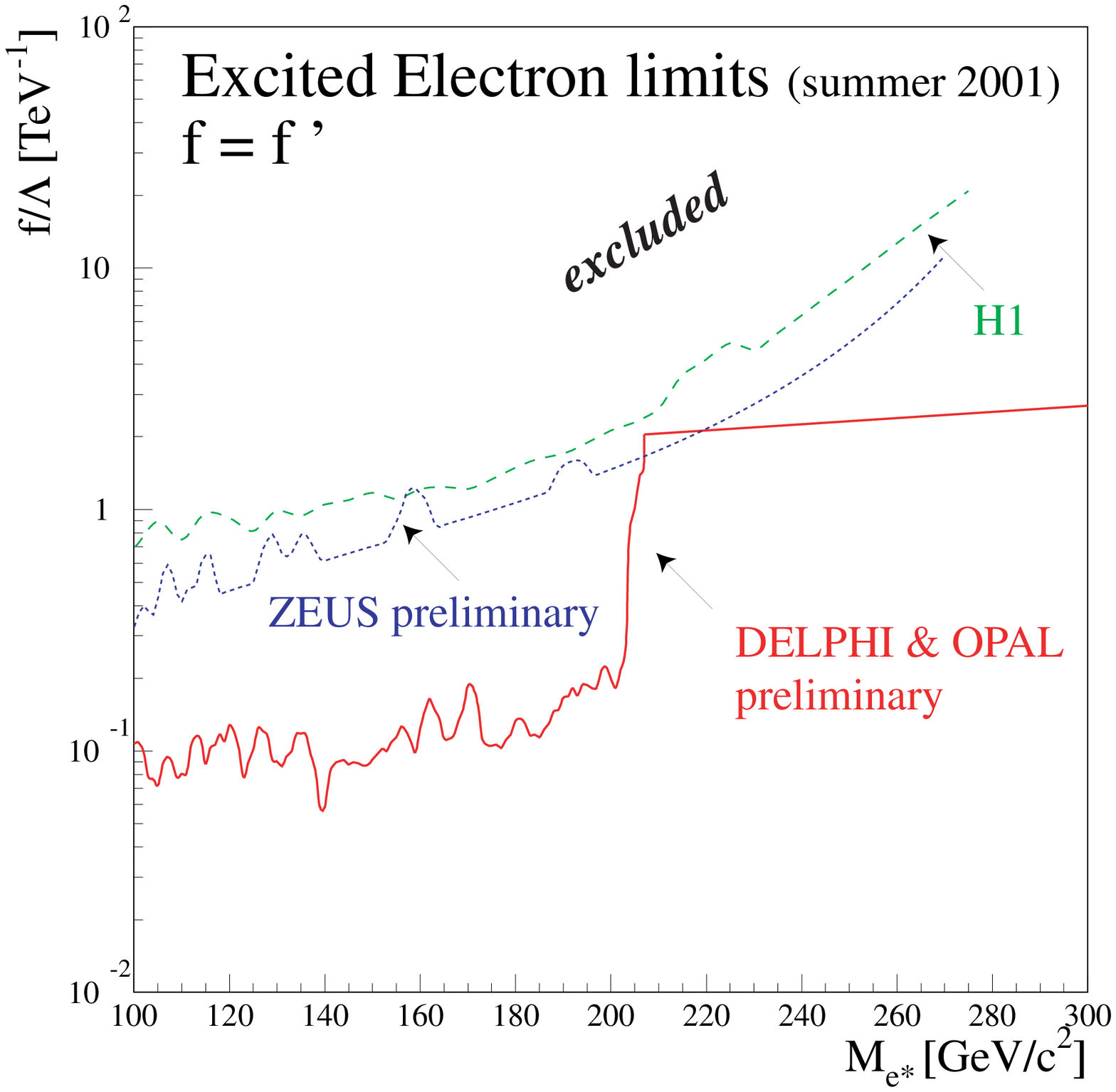,height=5.5cm}
\epsfig{figure=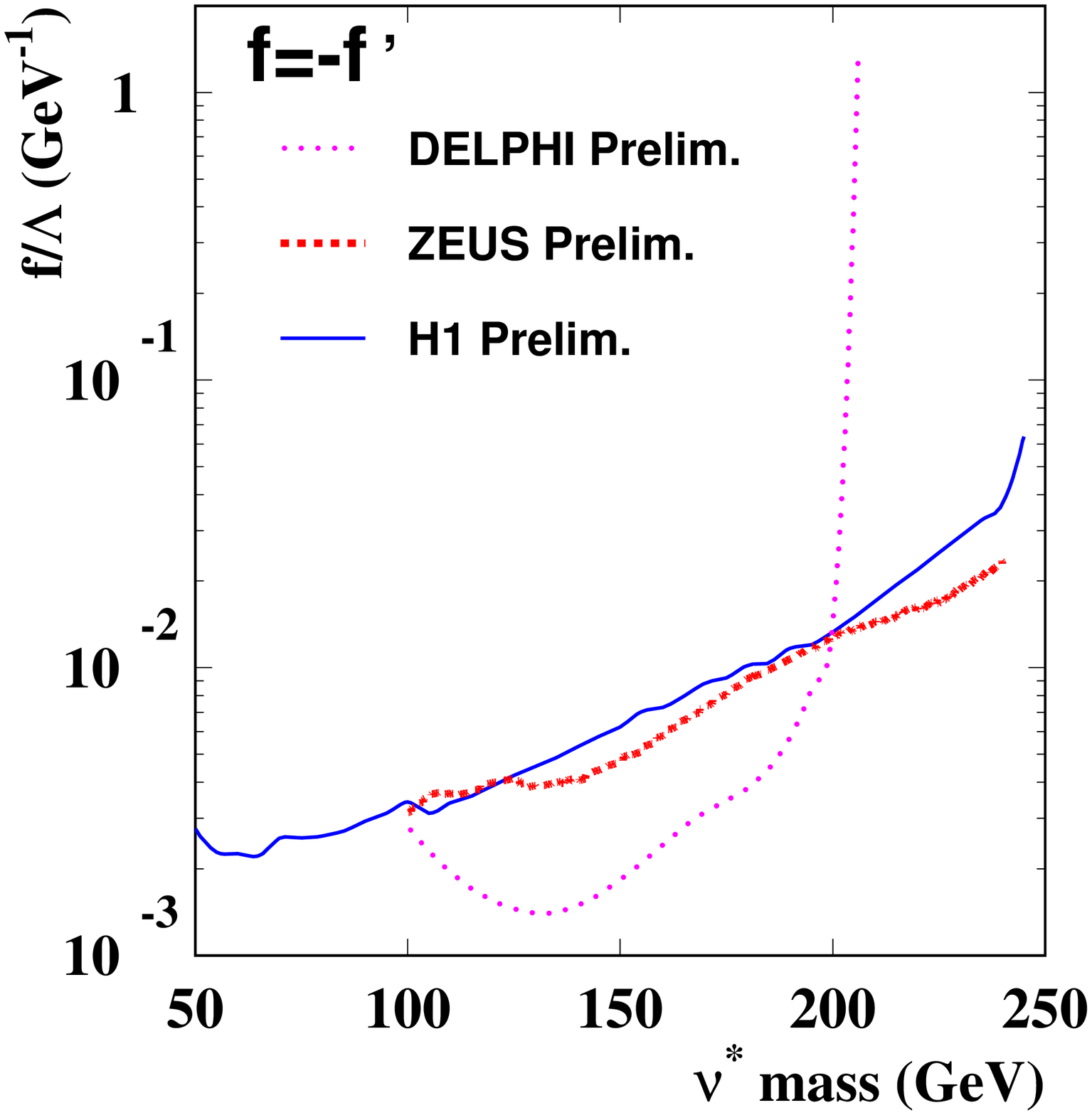,height=5.5cm}
\epsfig{figure=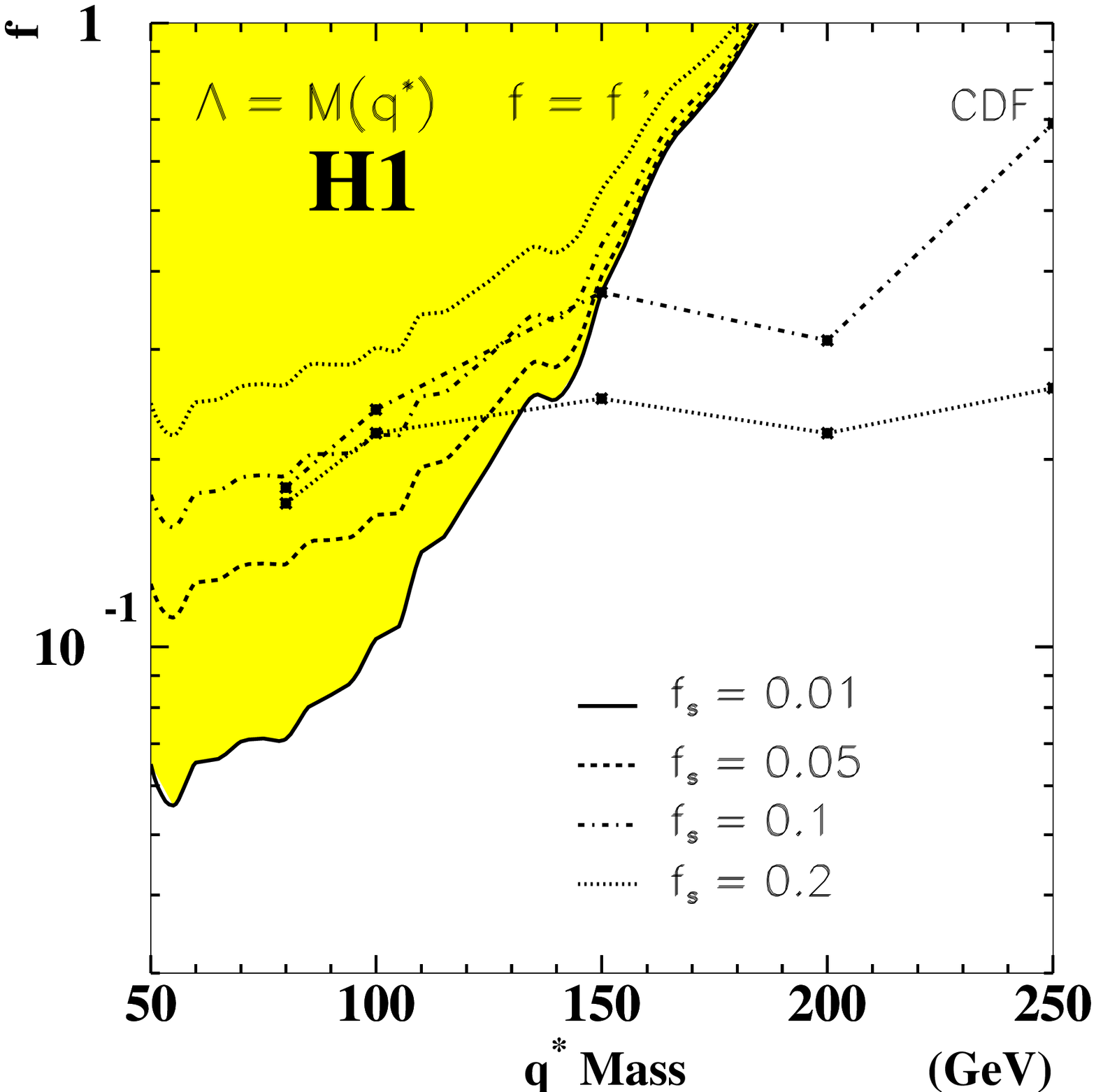,height=5.5cm}
$$
\vspace{-1.cm}
\caption{ \it Limits on $f/\lambda$ versus the 
$e^*$, $\nu^*$ and $q^*$ mass.
The analysed data are respectively 
98-00 $e^\pm p$ (ZEUS), 98-99 $e^-p$ (ZEUS and H1) and 94-97 $e^+p$ 
(H1).
In the $q^*$ channel, limits are set on $f$ by fixing $\Lambda$ to
the excited quark mass.
Tevatron limits are shown for different values of $f_s$.}
\label{fig:eflim}
\end{figure}

The resulting limits on $f/\Lambda$ for each channel are presented in
figure \ref{fig:eflim} as a function of the excited fermion mass
\cite{h1excited}. In the 
excited electron channel, HERA limits are competitive with LEP limits for
high excited electron masses. In the excited neutrino channel, HERA 
has a unique sensitivity at high excited neutrino masses. Finally, in
the excited quark channel, HERA limits are more sensitive than Tevatron limits
if small values of $f_s$ are considered.

\begin{figure}[htbp]
\vspace{-0.5cm}
$$
\epsfig{figure=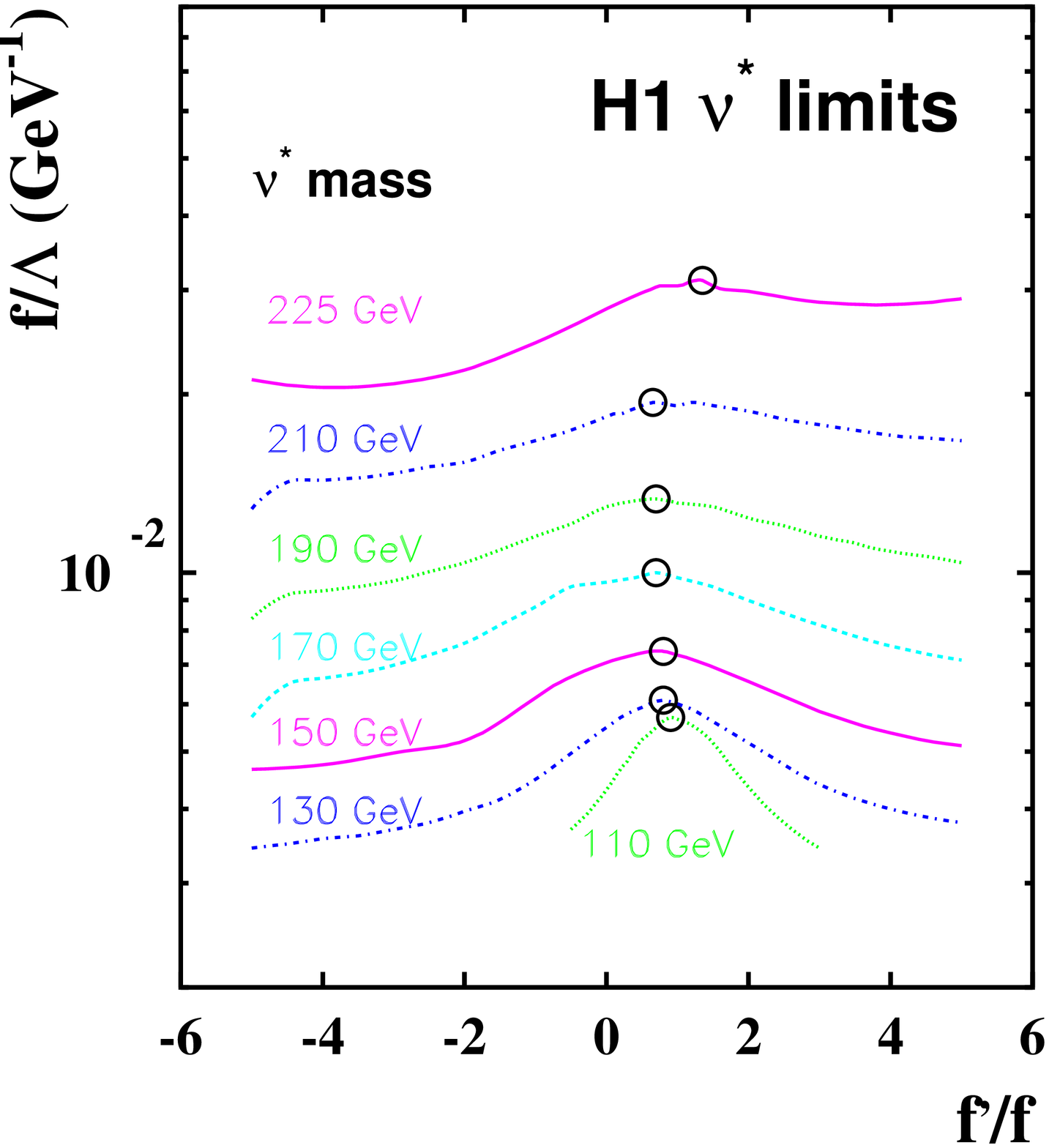,height=8cm}
\epsfig{figure=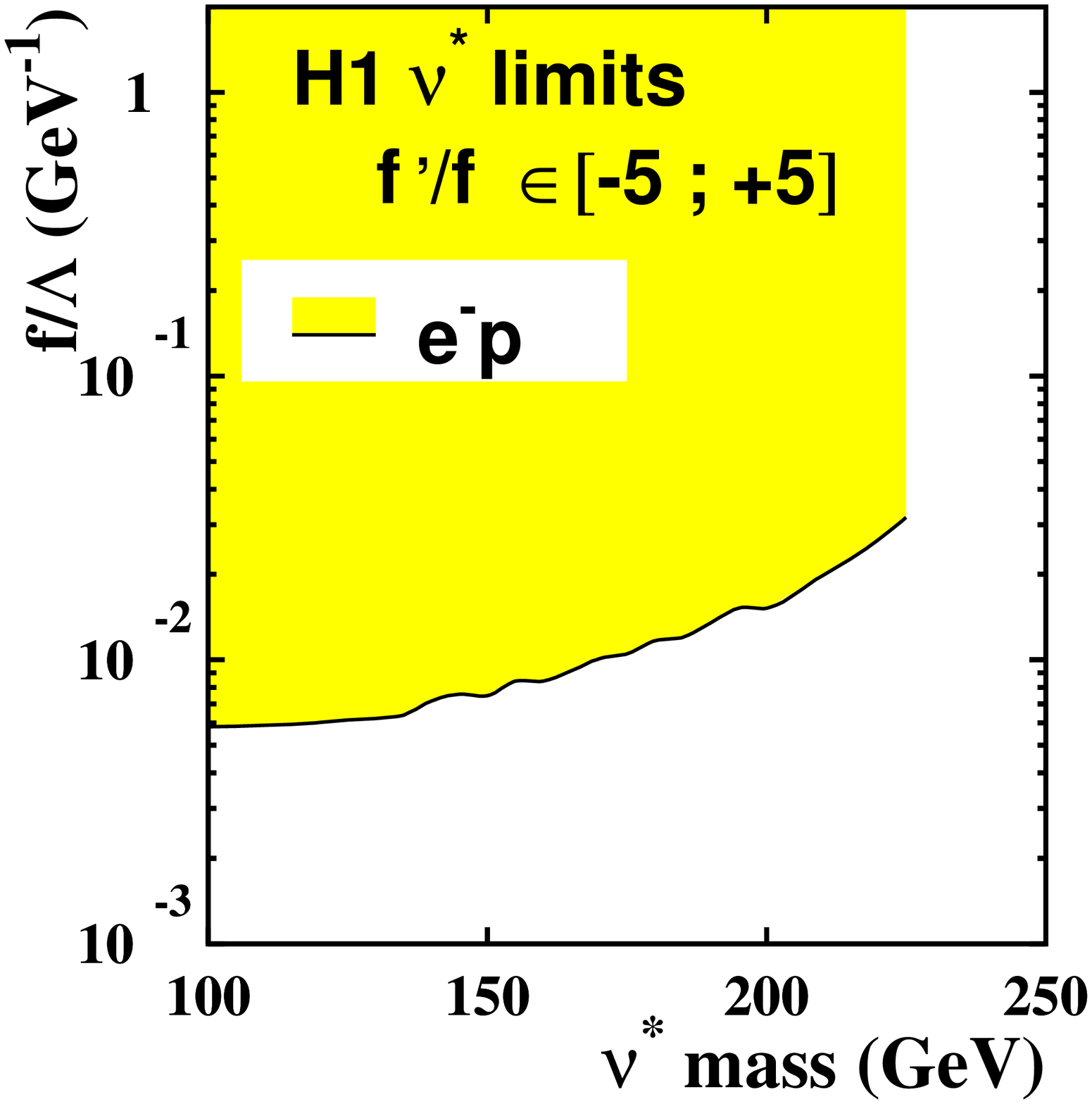,height=7.3cm}
$$
\vspace{-1.cm}
\caption{ \it The left plot presents the limits on $f/\Lambda$ as a 
function of $f'/f$
 for different values of the excited neutrino mass. The 
right plot shows the
resulting limits on $f/\Lambda$ as a function of the excited neutrino mass
for a range of $f'/f \in [-5;+5]$.}
\label{fig:efend}
\end{figure}

Less model dependent limits are also obtained by allowing an arbitrary 
ratio $f'/f$ in the excited neutrino analysis. Figure \ref{fig:efend}
shows the limits on $f/\Lambda$ as a function of the ratio $f'/f$ for different
values of the excited neutrino mass. These limits are then converted into 
limits versus the excited neutrino mass for a range of $f'/f$ between $-5$
and $5$.

\section{Isolated leptons}

\begin{figure}[htbp]
\vspace{-3.3cm}
$$
\hspace{4.5cm}
\epsfig{figure=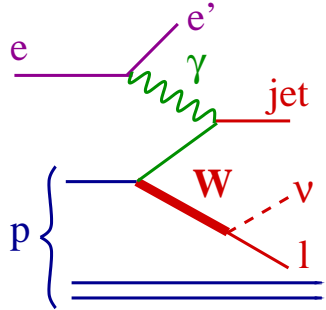,height=3.5cm}
$$
\vspace{3.6cm}
\caption{ \it W production at HERA}
\label{fig:wdiag}
\end{figure}

High $P_T$ isolated lepton events with missing $P_T$
were observed in the 1994-97 H1 data
\cite{observation}.
This analysis has now been extended to the full HERA-1 data. The most likely
SM interpretation of these events is W production with a lepton
decay of the W as illustrated in figure \ref{fig:wdiag}.

The LO cross section of the W production has been calculated and evaluated
to be at the level of 1.3 pb. NLO calculations reduce the scale uncertainty to about
$10~\%$ at high $P_T^X$ \cite{schwan}.

W production events are characterized by missing $P_T$ (due to the presence of the
neutrino), a high energy lepton and possibly a jet. The electron and muon 
channels of the W decay ($W \rightarrow e \nu$ and $W \rightarrow \mu \nu$)
 have been analysed by H1 and ZEUS.

\begin{figure}[htbp]
$$
\epsfig{figure=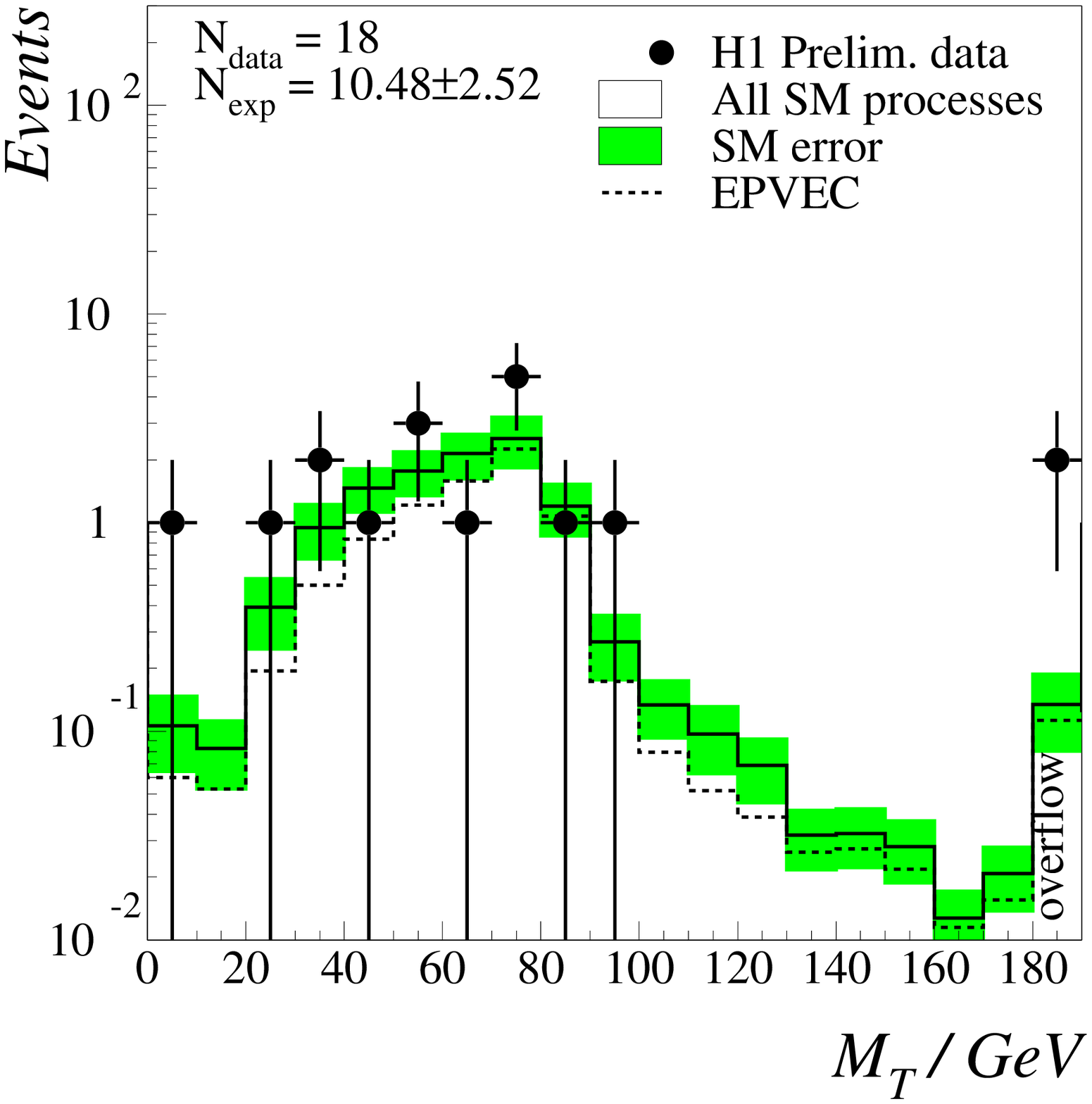,height=5.cm}
\epsfig{figure=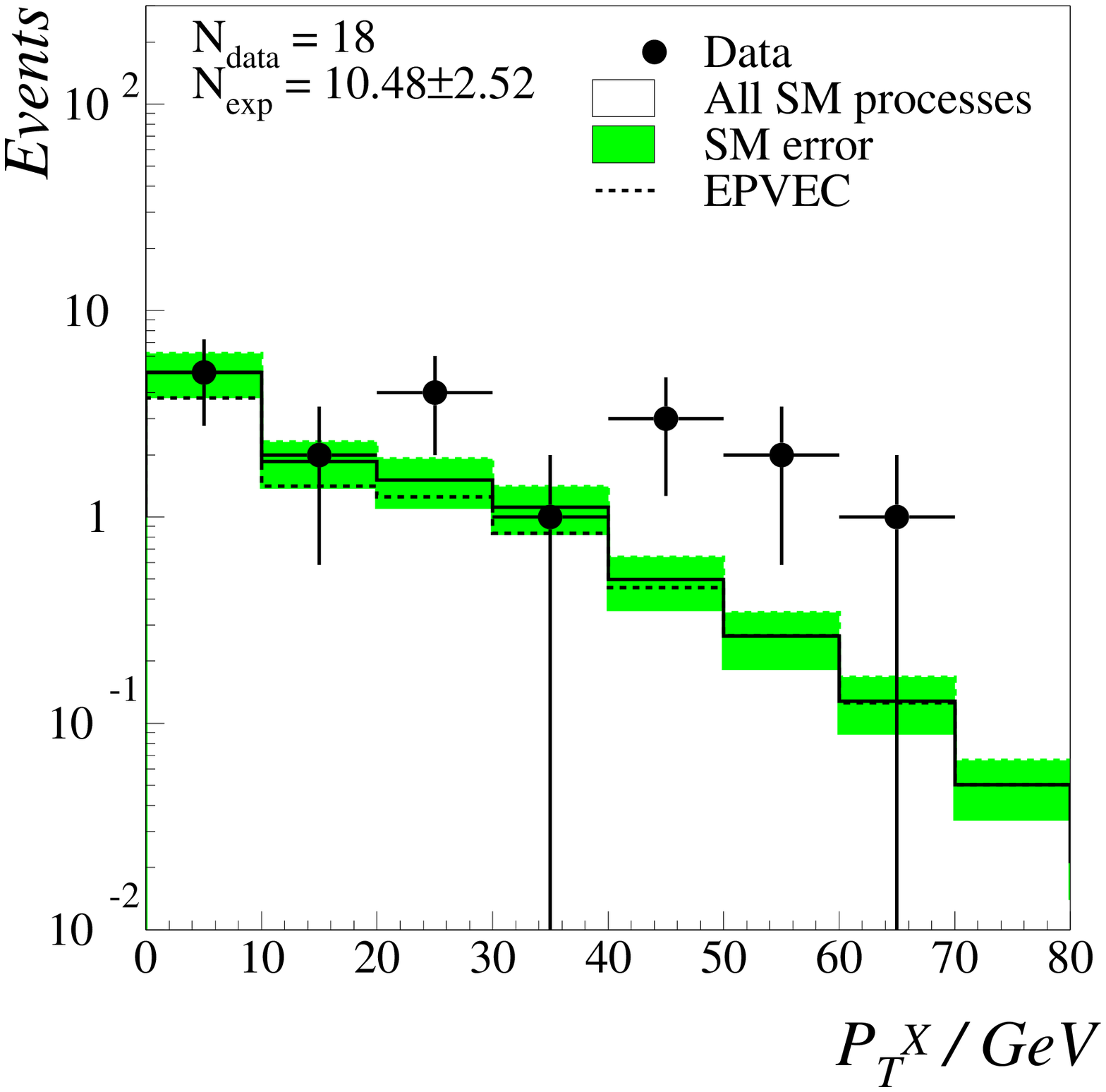,height=5.cm}
$$
\vspace{0.5cm}
\caption{ \it Distribution of the lepton-neutrino 
transverse mass $M_T$ and the 
hadronic system transverse momentum $P_T^X$ of the H1 data.}
\label{fig:kin}
\end{figure}

Table \ref{tab:lepton} shows the H1 and ZEUS event rates with the full
HERA-1 statistics in both channels
as a function of the transverse momentum of the hadronic system $P_T^X$
\cite{h1isol}.
The distribution of the lepton-neutrino transverse mass and the 
hadronic system transverse momentum $P_T^X$ are presented 
in figure \ref{fig:kin}.
At low $P_T^X$ the observed events are consistent with the SM
expectation. However more events than expected are observed by H1 at high
$P_T^X$.

\bf
\begin{table}[htbp] 
\begin{center}
\begin{tabular}{|c|c|c|}
\hline
Data/SM & H1 Prelim. & ZEUS Prelim. \\
& ($e^+p,~102~\mbox{pb}^{-1}$) & ($e^\pm p,~130~\mbox{pb}^{-1}$) \\
\hline
$~P_T^X>0~GeV~$ & $18/10.5 \pm 2.5$ & $-$ \\
\hline
$~P_T^X>25~GeV~$ & $10/2.8 \pm 0.7$ & $2/2.4 \pm 0.2$ \\
\hline
$~P_T^X>40~GeV~$ & $6/1.0 \pm 0.3$ & $0/1.0 \pm 0.1$ \\
\hline
\end{tabular}

\begin{tabular}{|c|}
\hline
$~$($e^-p,~14~\mbox{pb}^{-1}$) \\
$0/1.8 \pm 0.4$ \\
\hline
\end{tabular}
\end{center}
\vspace{-0.5cm}
\caption{ \it H1 and ZEUS results (number of data events / MC expectation)
in the combined electron 
and muon channels for the full HERA-1 data sample.}
\label{tab:lepton}
\end{table}
\rm

\begin{figure}[htbp]
$$
\epsfig{figure=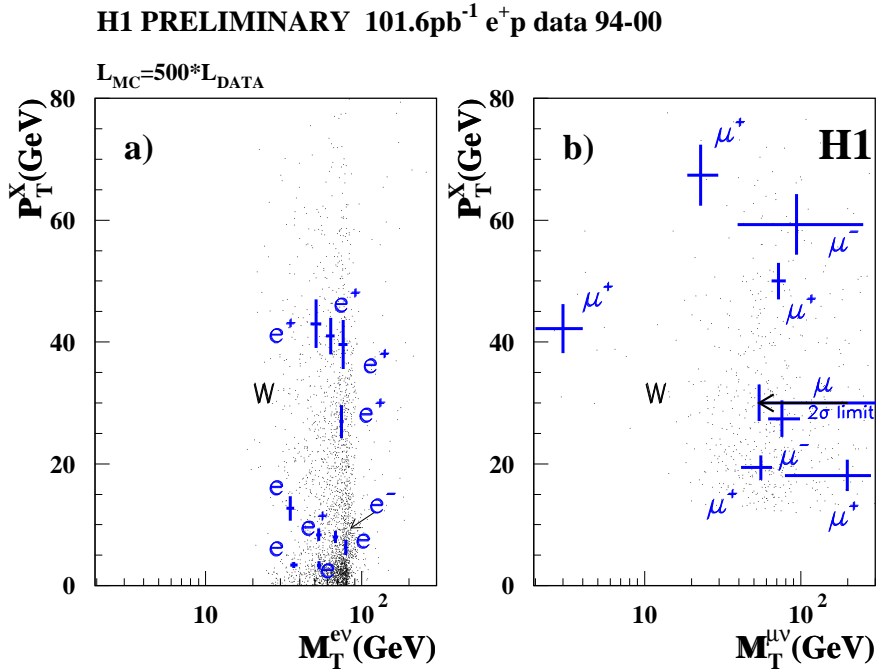,height=12.cm}
$$
\vspace{-3.5cm}
\caption{ \it Hadronic transverse momentum $P_T^X$ versus the lepton-neutrino
transverse mass for electron (left) and muon (right) channels.}
\label{fig:2d}
\end{figure}

Figure \ref{fig:2d} shows the 2-dimensional ($P_T^X$,$M_T$) distribution. 
Few data events are observed in regions where the 
SM expectation is low.
A possible interpretation of these events is the anomalous production
of single top events.

\section{Single top production}

The top quark production within the SM is negligible at HERA. Therefore
if any top signal would be found, this would be a sign of new physics.
The top quark could be produced by Flavour Changing Neutral Current 
(FCNC) through a $\gamma u$ fusion \cite{fcnc}.
In the semi-leptonic decay channel ($t\rightarrow bW$ with $W\rightarrow l\nu$)
the final state is composed of an isolated lepton, missing transverse momentum
and high hadronic transverse momentum $P_T^X$.

\begin{figure}[htbp]
$$
\epsfig{figure=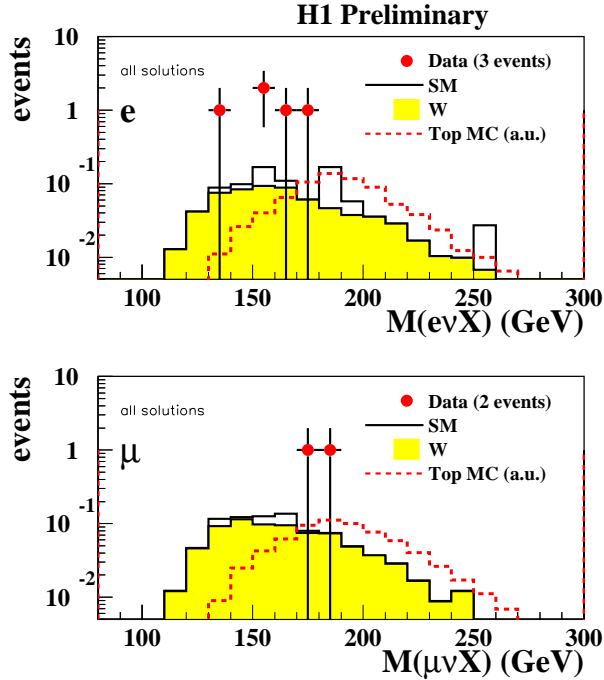,height=10.cm}
$$
\vspace{-1.cm}
\caption{ \it Mass distribution of the lepton-neutrino-jet 
system for electron (upper
plot) and muon (lower plot) channels. All mass solutions are considered
in the $\nu$ reconstruction.}
\label{fig:ltopmass}
\end{figure}

The W selection is further tightened in order to select
top candidates \cite{h1singletop,zeussingletop}. 
Figure \ref{fig:ltopmass} presents the resulting mass distribution of the 
lepton-neutrino-jet system for the electron and muon channels. Several
lepton events are in the top mass region.

A complementary top search has been performed via the hadronic channel
($t\rightarrow b W$ with $W\rightarrow q \overline{q}$) with three high $P_T$ jets
expected in the final state.

\begin{figure}[htbp]
\vspace{-4.2cm}
$$
\hspace{-3.cm}
\epsfig{figure=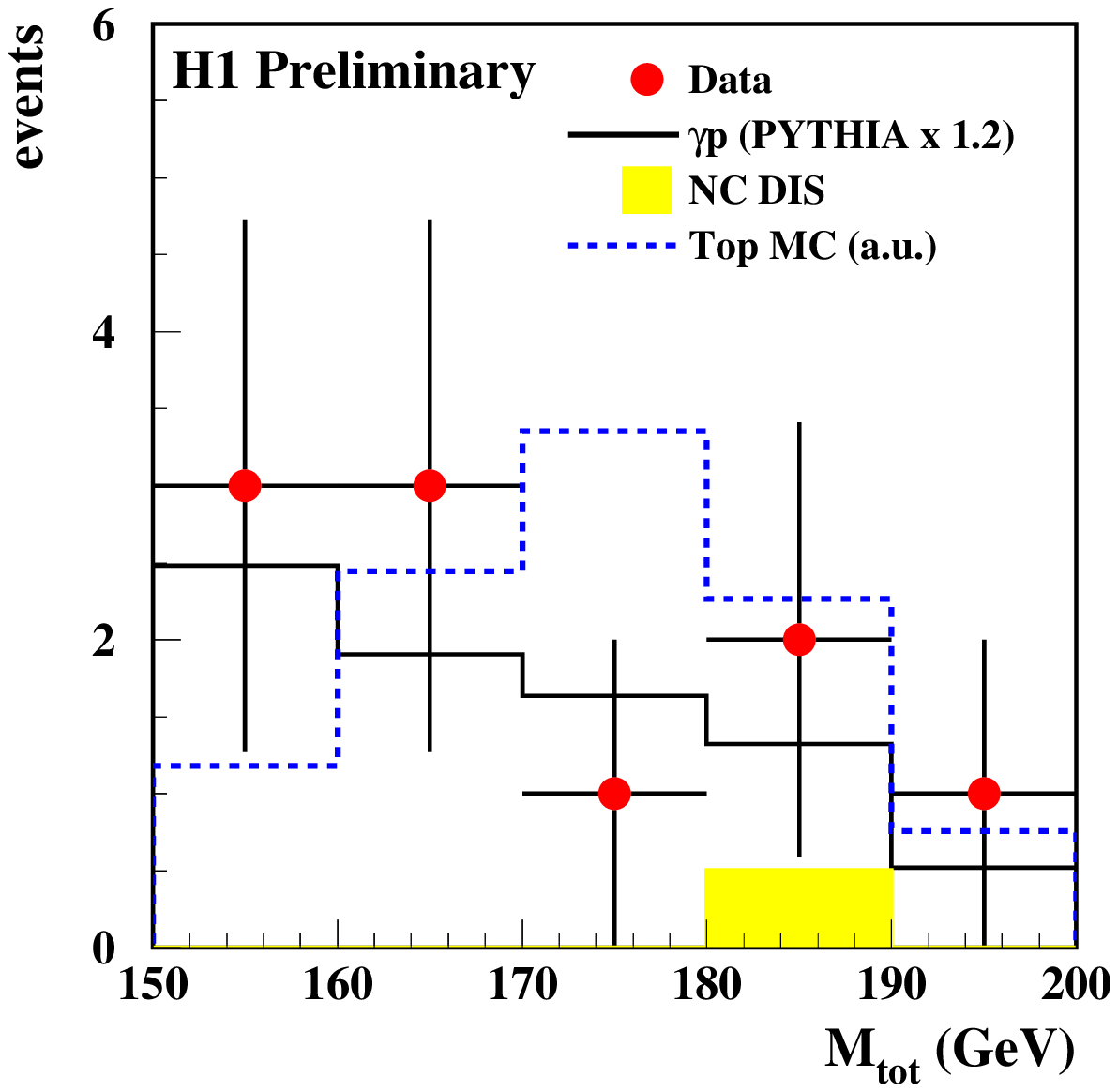,height=5.5cm}
$$
\vspace{-3.5cm}
$$
\hspace{4.cm}
\epsfig{figure=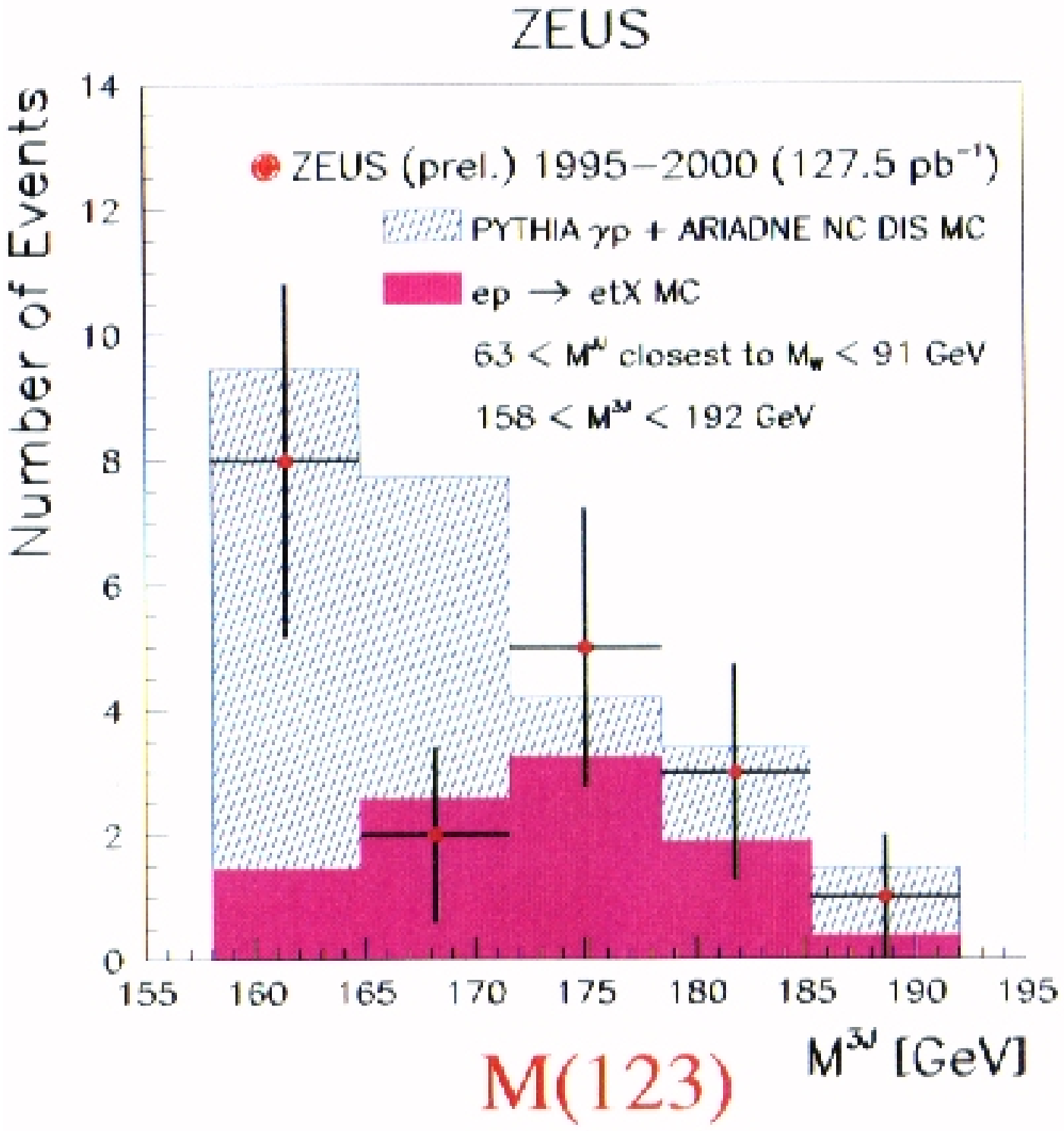,height=5.5cm}
$$
\vspace{-0.5cm}
\caption{ \it The left (right) figure shows the 3-jet mass distribution of the
selected top candidates of the H1 (ZEUS) experiment for a luminosity of 37 
$\mbox{pb}^{-1}$ (128 $\mbox{pb}^{-1}$).}
\label{fig:3jet}
\end{figure}

In addition to the existing analysis, further cuts are applied. A 2-jet mass
is reconstructed in a W mass window in order to reduce the signal/noise
ratio. A 3-jet mass is also reconstructed as shown in figure \ref{fig:3jet}.

The results of the top searches are shown in table \ref{tab:ltop} for both
semi-leptonic and hadronic decay channels.
No visible excess is observed in the hadronic channel compared to the SM,
therefore limits on anomalous top couplings are derived.

\bf
\begin{table}[htbp] 
\begin{center}
\begin{tabular}{|c|c|c|}
\hline
Data/SM & Lepton channel & Hadron channel\\
\hline
H1 & $5/1.8 \pm 0.5$  & $10/8.3^{+4.2}_{-1.9}\pm 4.2$\\
\hline
ZEUS & $0/1.0$ & $19/20.0$\\
\hline
\end{tabular}
\end{center}
\vspace{-0.5cm}
\caption{ \it Single top results (number of data events / MC expectation) 
in leptonic channel 
with a luminosity of 115 (128) $\mbox{pb}^{-1}$ for H1 (ZEUS),
and hadronic channel
with a luminosity of 37 (128) $\mbox{pb}^{-1}$ for H1 (ZEUS).}
\label{tab:ltop}
\end{table}
\rm

\newpage
HERA resulting limits on $tu\gamma$ coupling are the following:

\begin{itemize}
\item H1: $\kappa_{tu\gamma} < 0.305$
\vspace{-0.2cm}
\item ZEUS: $\kappa_{tu\gamma} < 0.19$
\end{itemize}

\begin{figure}[hhhh]
$$
\epsfig{figure=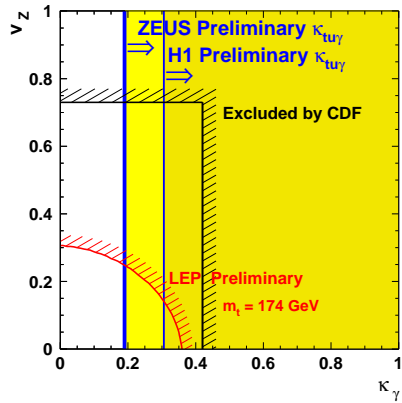,height=6.cm}
$$
\vspace{-1.2cm}
\caption{ \it Exclusion region in the plane ($v_Z$,$\kappa_\gamma$).}
\label{fig:final}
\end{figure}

Figure \ref{fig:final} summarizes the limits on the anomalous coulings 
$V_Z$ ($tuZ$ vectorial coupling) and $\kappa_\gamma$ ($tu\gamma$
magnetic coupling) obtained at HERA, LEP and Tevatron.
HERA sensitivity to $\kappa_\gamma$ is competitive with other colliders.

\section{Conclusion and future prospects}

HERA is the unique collider to test direct $eq$ interactions. About
$110~\mbox{pb}^{-1}$ of $e^+p$ data and $15~\mbox{pb}^{-1}$ of $e^-p$ data
have been collected per experiment at HERA-1.

No evidence of new physics has been observed in various models in inclusive
analyses (contact interactions, extra-dimensions, leptoquarks)
and exclusive analyses (lepton-flavour
violation, $R_p$-violating SUSY, excited fermions),
therefore new constraints have
been set.
HERA limits are seen to be competitive with and complementary to the LEP and 
Tevatron searches. The status of isolated lepton events 
with missing $P_T$ is still intriguing
and will become clearer with the new HERA-2 data. 

HERA has been shutdown since fall 2000 for a general upgrade:
new focussing magnets have been installed 
in order to increase the luminosity and many improvements have been performed in 
the detectors in order to increase their sensitivity.
Moreover, the lepton beam will be longitudinally
polarised in the H1 and ZEUS interaction regions.
The first luminosity
runs are predicted for beginning 2002 and HERA-2 is expected to accumulate 1
$\mbox{fb}^{-1}$ in the next 5 years. 
The anticipated
factor of ten increase in the integrated luminosity will give an
outstanding discovery potential for HERA.

\section{Acknowledgements}
I which to thank my H1 and ZEUS colleagues who contributed to the results 
presented here as well as people who helped me in preparing this talk.

\end{document}